\documentclass[
twocolumn,
pre,
aps,
superscriptaddress,
showpacs,
preprintnumbers,
amsmath,
amssymb]{revtex4-1}
\usepackage{bm}

\usepackage[dvipdfmx]{graphicx}
\usepackage[dvipdfmx,usenames]{color}
\begin{document}

\title{The statistical physics of discovering exogenous and endogenous factors in a chain of events}

\author{Shinsuke Koyama}
\email{skoyama@ism.ac.jp}
\affiliation{The Institute of Statistical Mathematics, Tokyo 190-8562, Japan}
\author{Shigeru Shinomoto}
\email{shinomoto.shigeru.6e@kyoto-u.ac.jp}
\affiliation{Department of Physics, Kyoto University, Kyoto 606-8502, Japan}
\affiliation{Brain Information Communication Research Laboratory Group, ATR Institute International, Kyoto 619-0288, Japan}

\begin{abstract}
Event occurrence is not only subject to the environmental changes, but is also facilitated by the events that have occurred in a system. Here, we develop a method for estimating such extrinsic and intrinsic factors from a single series of event-occurrence times. The analysis is performed using a model that combines the inhomogeneous Poisson process and the Hawkes process, which represent exogenous fluctuations and endogenous chain-reaction mechanisms, respectively. The model is fit to a given dataset by minimizing the free energy, for which statistical physics and a path-integral method are utilized. Because the process of event occurrence is stochastic, parameter estimation is inevitably accompanied by errors, and it can ultimately occur that exogenous and endogenous factors cannot be captured even with the best estimator. We obtained four regimes categorized according to whether respective factors are detected. By applying the analytical method to real time series of debate in a social-networking service, we have observed that the estimated exogenous and endogenous factors are close to the first comments and the follow-up comments, respectively. This method is general and applicable to a variety of data, and we have provided an application program, by which anyone can analyze any series of event times. 
\end{abstract}

\maketitle

%%%%%%%%%%%%%%%%%%%%%%%%%%%%%%%%%%%%
\section{Introduction}
%%%%%%%%%%%%%%%%%%%%%%%%%%%%%%%%%%%%

It is widely seen that events occur one after another in a chain by self-excitation mechanisms, as in the communication of diseases~\cite{Hethcote2000, Keeling2011, Vespignani2012, Brockmann2013, Pastorsatorras2015}, crimes and conflict~\cite{Goffman1964, Kitsak2010, Mohler2011, Lewis2012}, seismic dynamics \cite{ogata1988, helmstetter2003}, scientometrics \cite{michael2012}, tweets~\cite{CraneSornette2008, Lerman2010, Romero2011}, financial events~\cite{AtSahalia2010, Hardiman2013, bacry2015, hawkes2018}, and neuronal firing~\cite{Pernice2011, ReynaudBouret2014, Kass2018, GerhardDegerTruccolo2017}. Generally, event occurrence is subject not only to such intrinsic self-excitation mechanisms, but also environmental changes that take place independently of events that have occurred in a system. For instance, contagious diseases may spread from one individual to another, but the occurrence rate may also be subject to environmental factors such as temperature change~\cite{Lipp2002}. When considering how to facilitate or attenuate occurrence activity, we need to know the potential contributions of endogenous and exogenous factors~\cite{Menezes2004,lazer2018science, Omi2017}. When the government reviews possible intervention in the economy, it must be able to project the degree of the uncontrollable chain reaction causing economic fluctuations such as falling stock prices, and hopefully to calculate the possible impact of external intervention.

It has been suggested that the temporal profile of changes in the tweeting rate may provide information as to whether such changes have been caused endogenously or exogenously~\cite{CraneSornette2008}, but opinions have been divided over the issue as for whether gross activity contains enough information. Several works have addressed the estimation problem by analyzing precise time series of event occurrence using machine-learning algorithms~\cite{Lerman2010, Romero2011, zaman2014, cheng2014, Dow2013, petrovic2011rt, Aoki2016, Cattuto2007, Rybski2009}. Recently, we have developed an analytical method based on a generalized linear model (GLM) equipped with a self-exciting mechanism and analyzed a real time series of tweets to determine the contributions of endogenous and exogenous factors~\cite{Fujita2018}. One problem with the GLM is its strong nonlinearity, such that previous events multiply each other and extrinsic and intrinsic factors are mixed up as a result. 

In the present study, we consider analyzing a series of occurrence times via the ``inhomogeneous Hawkes process,'' a linear superposition of the inhomogeneous Poisson and Hawkes processes~\cite{hawkes1971}, which represent exogenous fluctuations and endogenous chain-reaction mechanisms, respectively. We devise the ``Hawkes decoder,'' a method of fitting the inhomogeneous Hawkes process to a given dataset. We also develop a path-integral method for computing free energy analytically. Given an event-generation process, the path-integral method can determine the parameters of the model representing the degree of an internal chain reaction and the impact of environmental changes. 

This estimation is inevitably accompanied by errors, because the event-generation process itself is stochastic. The estimation errors can be assessed by applying the decoder to synthetic data generated by simulating the inhomogeneous Hawkes process and comparing the estimated and original parameters. It may occur that exogenous fluctuation and/or endogenous self-excitation are not captured, even by a Hawkes decoder that can represent the original process. Using a path-integral method, we construct phase diagrams in which the results are categorized into four qualitatively different regimes according to whether or not their respective factors are detected. Then, we apply the Hawkes decoder to a time series of debate in a social-networking service (SNS) to estimate the relative contributions of exogenous and endogenous factors. We also examine the model's capability of predicting the amount of follow-up comments induced by first comments in the SNS. We also provide an application program, by which anyone can analyze any sequence of event times.

%%%%%%%%%%%%%%%%%%%%%%%%%%%%%%%%%%%%
\section{Methods}
%%%%%%%%%%%%%%%%%%%%%%%%%%%%%%%%%%%%

%%%%%%%%%%%%%%%%%%%%%%%%%%%%%%%%%%%%
\subsection{Generating a series of events with the inhomogeneous Hawkes process}
%%%%%%%%%%%%%%%%%%%%%%%%%%%%%%%%%%%%

\begin{figure}[htbp]
\begin{center}
\includegraphics[width=8.6cm]{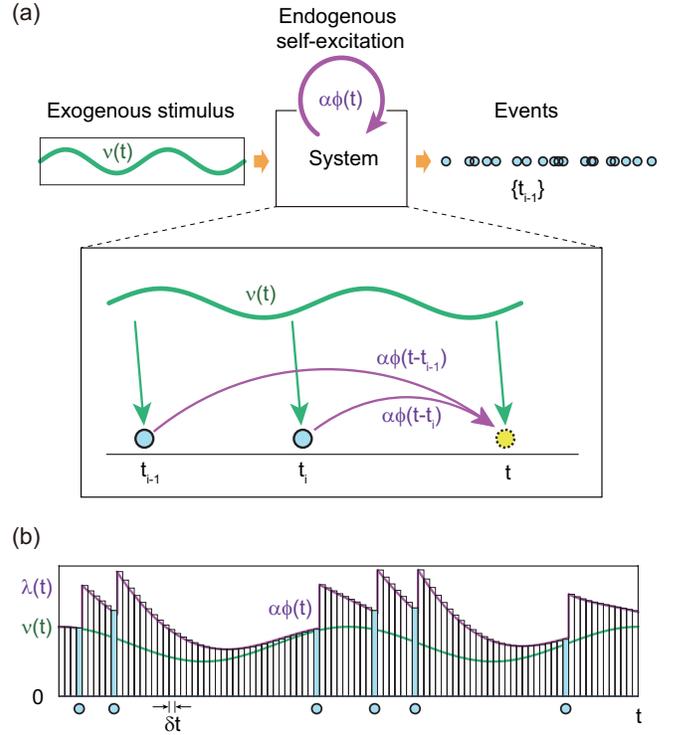}
\caption{
(a) Schematic representation of the inhomogeneous Hawkes process. 
Exogenous or extrinsic stimulus is represented by the time-dependent rate $\nu(t)$ of the inhomogeneous Poisson process, while endogenous or intrinsic self-excitation is represented by the (homogeneous) Hawkes process such that the occurrence of each event facilitates the occurrence of events by adding the occurrence rate $\alpha \phi (t)$. 
(b) A basic procedure for executing the inhomogeneous Hawkes process. Divide the time axis into small time intervals of width $\delta t$ and generate an event if a uniform random number $x \in [0,1]$ is smaller than $\lambda (t) \delta t$, and no event otherwise. 
}
\label{fig:schema}
\end{center}
\end{figure}

We define the inhomogeneous Hawkes process in terms of the intensity or the occurrence rate $\lambda(t)$ given by
\begin{equation}
\lambda(t) = \nu(t) + \alpha\sum_{t_i < t}\phi(t-t_i),
\label{eq:hawkes}
\end{equation}
where the first term $\nu(t)$ on the right-hand-side represents the inhomogeneous Poisson process such that events are derived from a time-varying occurrence rate given exogenously, whereas the second term represents a self-excitation effect in terms of the Hawkes process such that the occurrence of each event modifies the probability of future events endogenously (Fig.~\ref{fig:schema} (a)). Here, $\alpha$ is the reproduction ratio representing the average number of events induced by a single event, $t_i$ is the occurrence time of a past ($i$th) event, and $\phi(t)$ is a kernel representing the time course of the self-excitation, satisfying the causality $\phi(t)=0$ for $t<0$ and the normalization $\int_0^{\infty}\phi(t)dt=1$. 

%%%%%%%%%%%%%%%%%%%%%%%%%%%%%%%%%%%%%%%%%%%%%%%%%%%%%%%%%%
%%%%%%%%%%%%%%%%%%%%%%%%%%%%%%%%%%%%%%%%%%%%%%%%%%%%%%%%%%

A basic procedure for putting the rate-model process into practice is to divide the time axis into small time intervals of $\delta t$, and then to repeat the Bernoulli process in which an event is either present (with a probability of $\lambda(t) \delta t \ll 1$) or absent (with a probability of $1- \lambda(t) \delta t$) at each time bin (Fig.~\ref{fig:schema}(b)). The probability of having no event for the first $N$ intervals and finally having an event at the $N+1$st interval is given by $\prod_{i=1}^{N} (1-\lambda(i \delta t ) \delta t) \lambda(n \delta t ) \delta t$. By taking the limit of the infinitesimal interval, the probability density of having an inter-event interval $t \equiv n \delta t$ is given by
\begin{equation}
\lim_{\delta t \to 0} \lambda(t) \prod_{i=1}^{t/\delta t}\left(1-\lambda (i \delta t) \delta t \right) 
= \lambda(t) \exp{\left(-\int_{0}^{t} \lambda (t') dt' \right)}.
\label{eq:discrete}
\end{equation}
Accordingly, the probability that events occur at times $\{t_i\} \equiv \{t_1,\ldots,t_n\}$ in a period $[0,T]$ is obtained as \cite{Snyder1991,Daley2002}
\begin{equation}
p (\{t_i\}) = \left(\prod_{i=1}^n\lambda(t_i)\right)\exp\left(-\int_0^T\lambda(t)dt\right).
\label{eq:pdf_hawkes}
\end{equation}

%%%%%%%%%%%%%%%%%%%%%%%%%%%%%%%%%%%%
\subsection{The Hawkes decoder: a method of estimating endogenous and exogenous factors}
%%%%%%%%%%%%%%%%%%%%%%%%%%%%%%%%%%%%

We wish to estimate the exogenous stimulus $\nu(t)$ and the degree of self-excitation $\alpha$ that have contributed to generating a given series of events that occurred at times $\{t_i\}$ in a period of $[0, T]$. The estimation can be done by the Empirical Bayes method, in which the parameter and hyperparameter are determined by maximizing the marginal likelihood as follows.

Here, we assume that the external input $\nu(t)$ is modulated slowly in time. This may be represented by a prior distribution that penalizes the large gradient:
\begin{equation}
p_{\gamma}(\{\nu(t)\}) = \frac{1}{Z(\gamma)}\exp\left( -\frac{1}{2\gamma^2}\int_0^T\left(\frac{d\nu(t)}{dt}\right)^2dt \right),
\label{eq:prior}
\end{equation}
where $\gamma$ is a hyperparameter that specifies the smoothness or flatness of $\nu(t)$ and $Z(\gamma)$ is the normalization constant
\begin{equation}
Z(\gamma) = \int \exp\left( -\frac{1}{2\gamma^2}\int_0^T\left(\frac{d\nu(t)}{dt}\right)^2dt \right) D\{\nu(t)\},
\end{equation}
where $D\{\nu(t)\}$ represents functional integration over all possible paths of external inputs $\nu(t)$. Given a time series of events $\{t_i\}$, the posterior distribution of $\{\nu(t)\}$ is obtained through Bayes' theorem as
\begin{eqnarray}
p_{\alpha,\gamma}(\{\nu(t)\}|\{t_i\}) &=&
\frac{p_{\alpha}(\{t_i\} | \{\nu(t)\})p_{\gamma}(\{\nu(t)\})}{p_{\alpha,\gamma}(\{t_i\})},
\label{eq:posterior}
\end{eqnarray} 
where $p_{\alpha}(\{t_i\} | \{\nu(t)\})$ is the conditional probability that events occur at times $\{t_i\}$, which is given by Eq.~(\ref{eq:pdf_hawkes}). Here we have denoted the self-exciting parameter $\alpha$ and the external input $\nu (t)$ explicitly as conditions, because they are used to specify the underlying rate $\lambda(t)$ in Eq.~(\ref{eq:hawkes}). 

The denominator $p_{\alpha,\gamma}(\{t_i\})$ is the marginal likelihood obtained by the marginalization integral:
\begin{equation}
p_{\alpha,\gamma}(\{t_i\}) = 
\int p_{\alpha}(\{t_i\} | \{\nu(t)\})p_{\gamma}(\{\nu(t)\})D\{\nu(t)\}.
\label{eq:marginal_likelihood}
\end{equation}

The parameter $\alpha$ and hyperparameter $\gamma$ are determined by maximizing the marginal likelihood:
\begin{equation}
\{ \hat{\alpha},\hat{\gamma} \} = \arg\max_{\alpha, \gamma}p_{\alpha,\gamma}(\{t_i\}).
\label{eq:estimate_hyperparameters}
\end{equation}
With the selected parameter and hyperparameter, the likely external input is obtained as the maximum \textit{a posteriori} (MAP) estimate, such that the posterior distribution (\ref{eq:posterior}) is maximized:
\begin{equation}
\{\hat{\nu}(t)\} = \arg\max_{\{\nu(t)\}}p_{\hat{\alpha},\hat{\gamma}}(\{\nu(t)\}|\{t_i\}).
\label{eq:estimate_rate}
\end{equation}
With the estimated $\hat{\alpha}$, $\{\hat{\nu}(t)\}$, and the given series of occurrence times $\{t_i\}$, 
we can estimate the underlying rate as 
\begin{equation}
\hat{\lambda}(t) = \hat{\nu}(t) + \hat{\alpha}\sum_{t_i < t}\phi(t-t_i).
\label{eq:conditional_intensity_estimate}
\end{equation}
The estimated exogenous rate $\{\hat{\nu}(t)\}$ depends largely upon the selected hyperparameter $\hat{\gamma}$. With a large $\hat{\gamma}$, the estimated rate fluctuates largely in time. If the exogenous fluctuation is small, it may occur that the estimated $\hat{\gamma}$ vanishes, in which case the estimated rate $\hat{\nu}(t)$ becomes constant. Another parameter $\hat{\alpha}$ represents the estimated level of self-excitation, and it might also occur that the reproduction ratio $\hat{\alpha}$ vanishes. Thus, there are four regimes categorized according to the combination of whether $\hat{\gamma} =0$ (homogeneous) or $\hat{\gamma} \ne 0$ (inhomogeneous) and whether $\hat{\alpha} =0$ (self-excitation undetected) or $\hat{\alpha} \ne 0$ (self-exciting detected). The nicknames of the four regimes are given in Table~\ref{tbl:regime}. The method for selecting the parameter and hyperparameter $\{ \hat{\alpha},\hat{\gamma} \}$ and estimating a time-dependent exogenous stimulus $\hat{\nu}(t)$ is explained in full detail in Appendix \ref{appendix:numerical_method}. A ready-to-use version of the web application, the source code, and examplary data sets are available at our website: http://{}www.ton.scphys.kyoto-u.ac.jp/{}\%7Eshino/{}hawkes

%%%%%%%%%%%%%%%%%%%%%%%%%%%%%%%%%%%%%%%%%%%%%%%%%%%%%%%%%%
\begin{table}[h]
\begin{center}
\begin{tabular}{|c|c|c|}
\hline
$\hat{\alpha}$ & $\hat{\gamma}$ & interpretation \\ 
\hline
0 & 0 & \, ``Poisson'' \, \\
0 & \, finite \, & \, ``Exo'' \, \\
\, finite \, & 0 & \, ``Endo'' \, \\ 
\, finite \, & \, finite \, & \, ``Exo+Endo'' \, \\
\hline
\end{tabular}
\end{center}
\caption{
Four regimes categorized according to the selected parameter $\alpha$ and hyperparameter $\gamma$.
``Poisson'': Neither endogenous self-excitation nor exogenous fluctuation is detected, $\{\hat{\alpha}=0, \hat{\gamma}=0\}$, and accordingly the process is considered to be close to the homogeneous Poisson process.
``Exo'': Only temporal variation of the exogenous origin is detected, $\{\hat{\alpha}=0, \hat{\gamma} \ne 0\}$, and the process is close to the inhomogeneous Poisson process.
``Endo'': Only endogenous self-excitation is detected, $\{\hat{\alpha} \ne 0, \hat{\gamma} = 0\}$.
``Exo+Endo'': Both exogenous and endogenous factors are detected, $\{\hat{\alpha} \ne 0, \hat{\gamma} \ne 0\}$.
}
\label{tbl:regime} 
\end{table}
%%%%%%%%%%%%%%%%%%%%%%%%%%%%%%%%%%%%%%%%%%%%%%%%%%%%%%%%%%

%%%%%%%%%%%%%%%%%%%%%%%%%%%%%%%%%%%%
\subsection{path-integral estimation of the free energy}
%%%%%%%%%%%%%%%%%%%%%%%%%%%%%%%%%%%%

In the previous section, we showed the procedure for fitting the inhomogeneous Hawkes process to a given series of occurrence times. Nevertheless, the parameter and hyperparameter $\{\hat{\alpha}, \hat{\gamma}\}$ can be determined even without a series of events, if the event-generating process is given. In this section, we perform this estimation analytically using the path-integral method. 

The marginalization in Eq.~(\ref{eq:marginal_likelihood}) can be represented as a path-integral:
\begin{eqnarray}
p_{\alpha,\gamma}(\{t_i\}) = 
\frac{1}{Z(\gamma)}\int\exp(-S[\nu(t)])D\{\nu(t)\}.
\label{eq:pathintegral}
\end{eqnarray}
By representing the action functional $S[\nu(t)] = -\log (p_{\alpha}(\{t_i\} | \{\nu(t)\}) p_{\gamma}(\{\nu(t)\})) - \log Z(\gamma)$ as an integral,
\begin{equation}
S[\nu(t)] = \int_0^TL(\dot{\nu},\nu)dt,
\label{eq:action}
\end{equation} 
a classical orbit that satisfies the Euler-Lagrange equation
\begin{equation}
\frac{\partial}{\partial t}\left(\frac{\partial L}{\partial\dot{\nu}}\right) 
- \frac{\partial L}{\partial \nu} =0
\label{eq:Euler-Lagrange}
\end{equation}
is the MAP solution $\{\hat{\nu}(t)\}$, given by Eq.~(\ref{eq:estimate_rate}). 

The path-integral in Eq.~(\ref{eq:pathintegral}) can be performed analytically by expanding the action functional up to the quadratic order in the deviation from the classical orbit:
\begin{equation}
S[\nu(t)] \approx S[\hat{\nu}(t)] + \frac{1}{2}\delta^2S_{\hat{\nu}(t)}[\delta \nu(t)],
\end{equation}
where $\delta^2S_{\hat{\nu}(t)}[\delta \nu(t)] $ denotes an action integral of the deviation around the classical orbit $\hat{\nu}(t)$, defined as
\begin{eqnarray*}
\int_0^T\left(
\left.\frac{\partial^2L}{\partial\dot{\nu}^2}\right|_{\hat{\nu}(t)} \delta\dot{\nu}^2
+ \left. 2 \frac{\partial^2L}{\partial\dot{\nu}\partial\nu}\right|_{\hat{\nu}(t)} \delta\dot{\nu}\delta\nu
+ \left. \frac{\partial^2L}{\partial\nu^2}\right|_{\hat{\nu}(t)} \delta\nu^2
\right)dt.
\nonumber
\end{eqnarray*}
The marginal likelihood is obtained as
\begin{equation}
p_{\alpha,\gamma}(\{t_i\}) \approx 
R e^{-S[\hat{\nu}(t)]},
\label{eq:marginal_likelihood_approx}
\end{equation}
where 
\begin{equation}
R = \frac{1}{Z(\gamma)}\int e^{-\frac{1}{2}\delta^2S_{\hat{\nu}(t)}[\delta \nu(t)]} D\{\delta\nu(t)\} 
\label{eq:quantum_effect}
\end{equation}
represents a ``quantum effect.'' The ``free energy'' is the negative logarithm of the marginal likelihood:
\begin{eqnarray}
F(\alpha,\gamma) &=& -\frac{1}{T}\log p_{\alpha,\gamma}(\{t_i\}) \nonumber\\
&=&
-\frac{1}{T}\log R + \frac{1}{T}S[\hat{\nu}(t)].
\label{eq:free_energy_def}
\end{eqnarray}

By decomposing the original exogenous input into a mean and fluctuation and expanding the action integral up to the quadratic order of the fluctuation, the free energy for a long series of events can be obtained analytically in terms of the mean input and the autocorrelation of fluctuation. The derivation of the free energy and the obtained formula are summarized in Appendix \ref{appendix:derivation_free_energy}.

%%%%%%%%%%%%%%%%%%%%%%%%%%%%%%%%%%%%
\section{Results}
%%%%%%%%%%%%%%%%%%%%%%%%%%%%%%%%%%%%

%%%%%%%%%%%%%%%%%%%%%%%%%%%%%%%%%%%%
\subsection{Analyzing synthetic data}
%%%%%%%%%%%%%%%%%%%%%%%%%%%%%%%%%%%%

%%%%%%%%%%%%%%%%%%%%%%%%%%%%%%%%%%%%
\subsubsection*{Event-generation model}

\begin{figure}[htbp]
\begin{center}
\includegraphics[width=8.6cm]{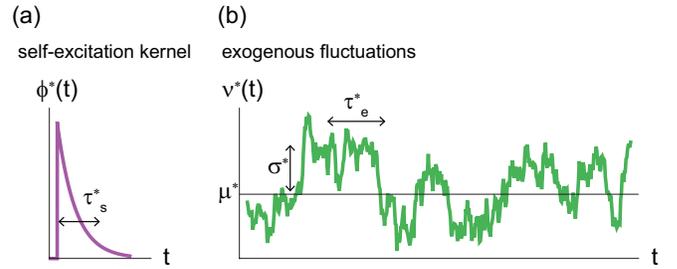}
\caption{Parameters of an inhomogeneous Hawkes process generating synthetic data.
(a) The self-excitation kernel $\phi^*(t) = 1/\tau^*_s \exp(-t/\tau^*_s)$. (b) Exogenous input $\nu^*(t)$ generated by the Ornstein-Uhlenbeck process (OUP) of given parameters $\mu^*$, $\sigma^*$, and $\tau^*_e$. }
\label{fig:synthetic}
\end{center}
\end{figure}

We shall examine the Hawkes decoder by applying it to a series of events generated by the inhomogeneous Hawkes process of a given reproduction ratio $\alpha^*$ and exogenous input $\{\nu^*(t)\}$:
\begin{equation}
\lambda^*(t) = \nu^*(t) + \alpha^* \sum_{t_i < t}\phi^*(t-t_i),
\label{eq:hawkes_original}
\end{equation}
where the self-excitation kernel is given as $\phi^*(t) = 1/\tau^*_s \exp(-t/\tau^*_s)$ for $t > 0$ and $\phi^*(t)=0$, otherwise. In particular, we examine the case where $\nu^*(t)$ is fluctuating according to the Ornstein-Uhlenbeck process (OUP) (Fig.~\ref{fig:synthetic}):
\begin{equation}
\frac{d\nu^*(t)}{dt} = -\frac{\nu^*(t)-\mu^*}{\tau^*_e} + \sigma^* \sqrt{\frac{2}{\tau_e^*}}\xi(t),
\label{eq:oup}
\end{equation}
where $\xi(t)$ is a Gaussian white noise satisfying $\langle \xi(t) \xi(t+u) \rangle = \delta (u)$. Accordingly, the OUP is characterized by an autocorrelation function with a time constant $\tau_e^*$:
\begin{equation}
\langle (\nu^*(t)-\mu^*) (\nu^*(t+u)-\mu^*) \rangle = {\sigma^*}^2 e^{-|u|/\tau_e^*}.
\label{eq:autocorrelation_oup}
\end{equation}
The inhomogeneous Poisson process with the fluctuating rate Eq.~(\ref{eq:oup}) is also called the doubly stochastic Poisson process. Thus, the current event-generation is a linear combination of the Hawkes process in Eq.~(\ref{eq:hawkes_original}) and the doubly stochastic Poisson process characterized by the autocorrelation Eq.~(\ref{eq:autocorrelation_oup}). The important dimensionless parameters of the model are the strength of endogenous self-excitation $\alpha^*$ and the exogenous fluctuation ${\sigma^*}^2 \tau_e^*/\mu^*$.

%%%%%%%%%%%%%%%%%%%%%%%%%%%%%%%%%%%%
%%%%%%%%%%%%%%%%%%%%%%%%%%%%%%%%%%%%%%%%%%%%%%%%%%%%%%%%%%

%%%%%%%%%%%%%%%%%%%%%%%%%%%%%%%%%%%%%%%%%%%%%%%%%%%%%%%%%%
\begin{figure*}[htbp]
\begin{center}
\includegraphics[width=12.9cm]{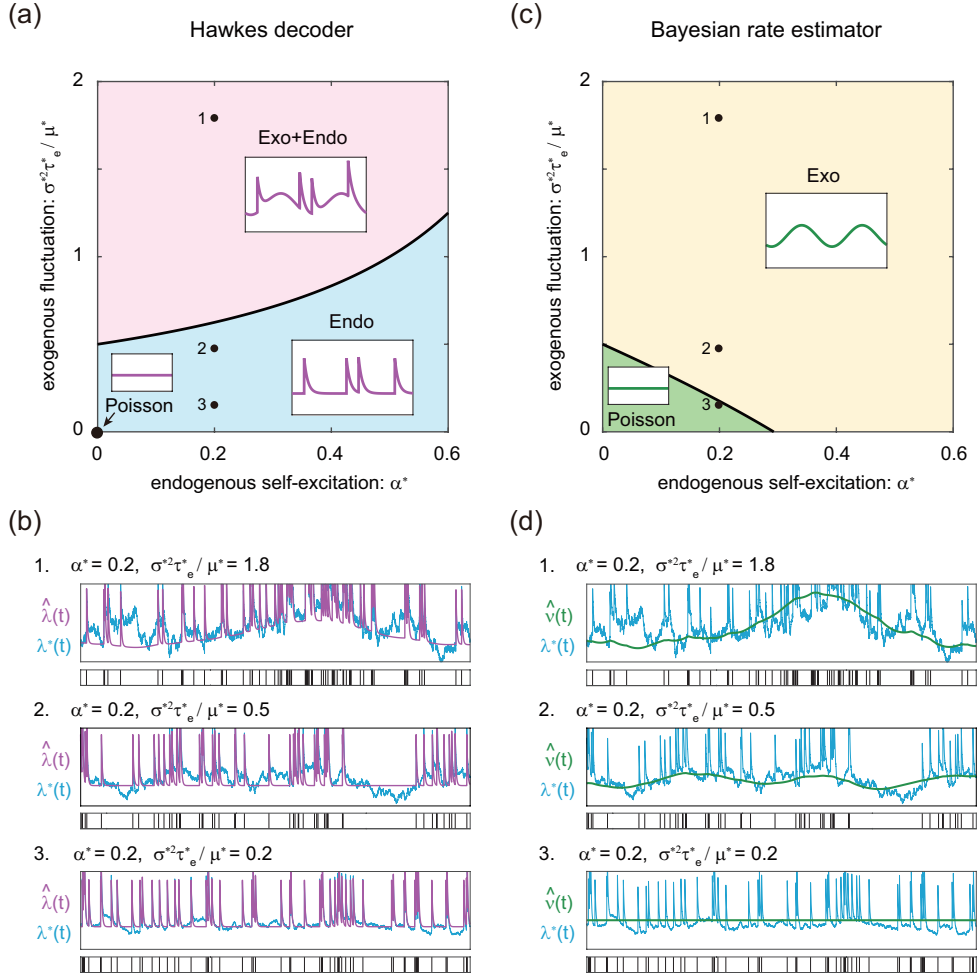}
\caption{
Phase diagrams of the ``Hawkes decoder'' and the ``Bayesian rate estimator'' for the doubly stochastic inhomogeneous Hawkes process characterized by an amplitude of the self-excitation $\alpha^*$ and exogenous fluctuation ${\sigma^*}^2 \tau_e^*/\mu^*$. (a) The Hawkes decoder. The parameters $\alpha$ and $\gamma$ are selected by minimizing the free energy $F(\alpha,\gamma)$. The ``Poisson,'' ``Endo,'' and ``Exo+Endo'' regimes are defined by $\{\hat{\alpha}= 0, \hat{\gamma}=0\}$, $\{\hat{\alpha}\ne 0, \hat{\gamma}=0\}$, and $\{\hat{\alpha}\ne 0, \hat{\gamma}\ne 0\}$, respectively. The phase boundaries are computed for the case of $\tau_e^* \gg \tau^*_s$. (b) The sample event series, the original rate $\lambda^*(t)$, and the rate $\hat{\lambda}(t)$ estimated by the Hawkes decoder. (c) The Bayesian rate estimator. The parameter $\alpha$ is chosen to be $0$ and the hyperparameter $\gamma$ is selected by minimizing the free energy $F_{\rm p}(\gamma) = F(\alpha=0,\gamma)$. The ``Exo'' regime is defined by $\{\hat{\alpha}=0, \hat{\gamma}\ne 0\}$. (d) The rate $\hat{\nu}(t)$ was estimated by the Bayesian rate estimator for the same event series analyzed in (b).
}
\label{fig:phasediagram_hawkes_poisson}
\end{center}
\end{figure*}
%%%%%%%%%%%%%%%%%%%%%%%%%%%%%%%%%%%%%%%%%%%%%%%%%%%%%%%%%%
%%%%%%%%%%%%%%%%%%%%%%%%%%%%%%%%%%%%%%%%%%%%%%%%%%%%%%%%%%
\begin{figure*}[htbp]
\begin{center}
\includegraphics[width=17.2cm]{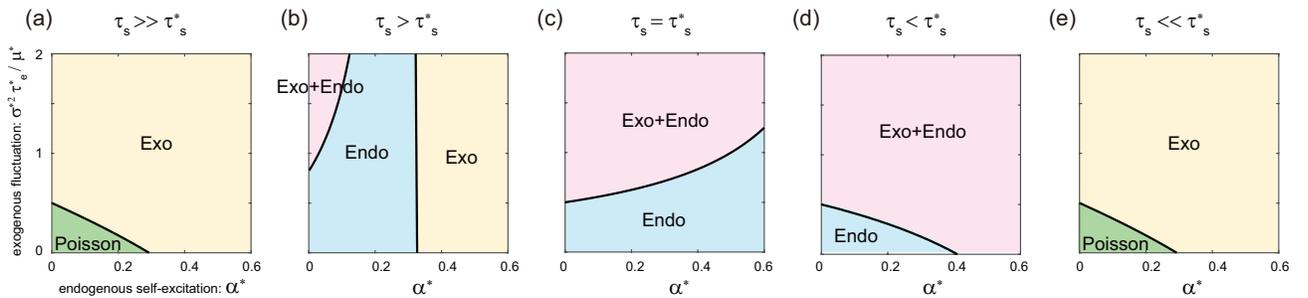}
\caption{
Phase diagrams of the Hawkes decoder assuming the kernels of various timescales. The timescale of the decoder's kernel $\tau_s$ is (a) $\tau_s \gg \tau_s^*$, (b) $\tau_s > \tau_s^*$, (c) $\tau_s = \tau_s^*$, (d) $\tau_s < \tau_s^*$, (e) $\tau_s \ll \tau_s^*$.
}
\label{fig:phasediagram_hawkes_wrong}
\end{center}
\end{figure*}
%%%%%%%%%%%%%%%%%%%%%%%%%%%%%%%%%%%%%%%%%%%%%%%%%%%%%%%%%%
%%%%%%%%%%%%%%%%%%%%%%%%%%%%%%%%%%%%%%%%%%%%%%%%%%%%%%%%%%
\begin{figure*}[htbp]
\begin{center}
\includegraphics[width=17.2cm]{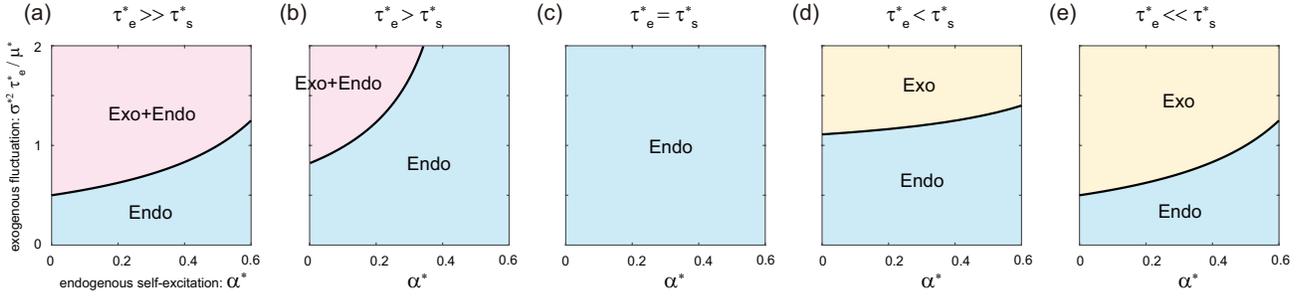}
\caption{
Phase diagrams of the Hawkes decoder with different timescales of the exogenous fluctuation $\tau_e$. The timescale of the exogenous fluctuation $\tau_e^*$ is (a) $\tau_e^* \gg \tau_s^*$, (b) $\tau_e^* > \tau_s^*$, (c) $\tau_e^* = \tau_s^*$, (d) $\tau_e^* < \tau_s^*$, (e) $\tau_e^* \ll \tau_s^*$.
}
\label{fig:phasediagram_hawkes_timescale}
\end{center}
\end{figure*}
%%%%%%%%%%%%%%%%%%%%%%%%%%%%%%%%%%%%%%%%%%%%%%%%%%%%%%%%%%
%%%%%%%%%%%%%%%%%%%%%%%%%%%%%%%%%%%%%%%%%%%%%%%%%%%%%%%%%%
\begin{figure*}[htbp]
\begin{center}
\includegraphics[width=12.9cm]{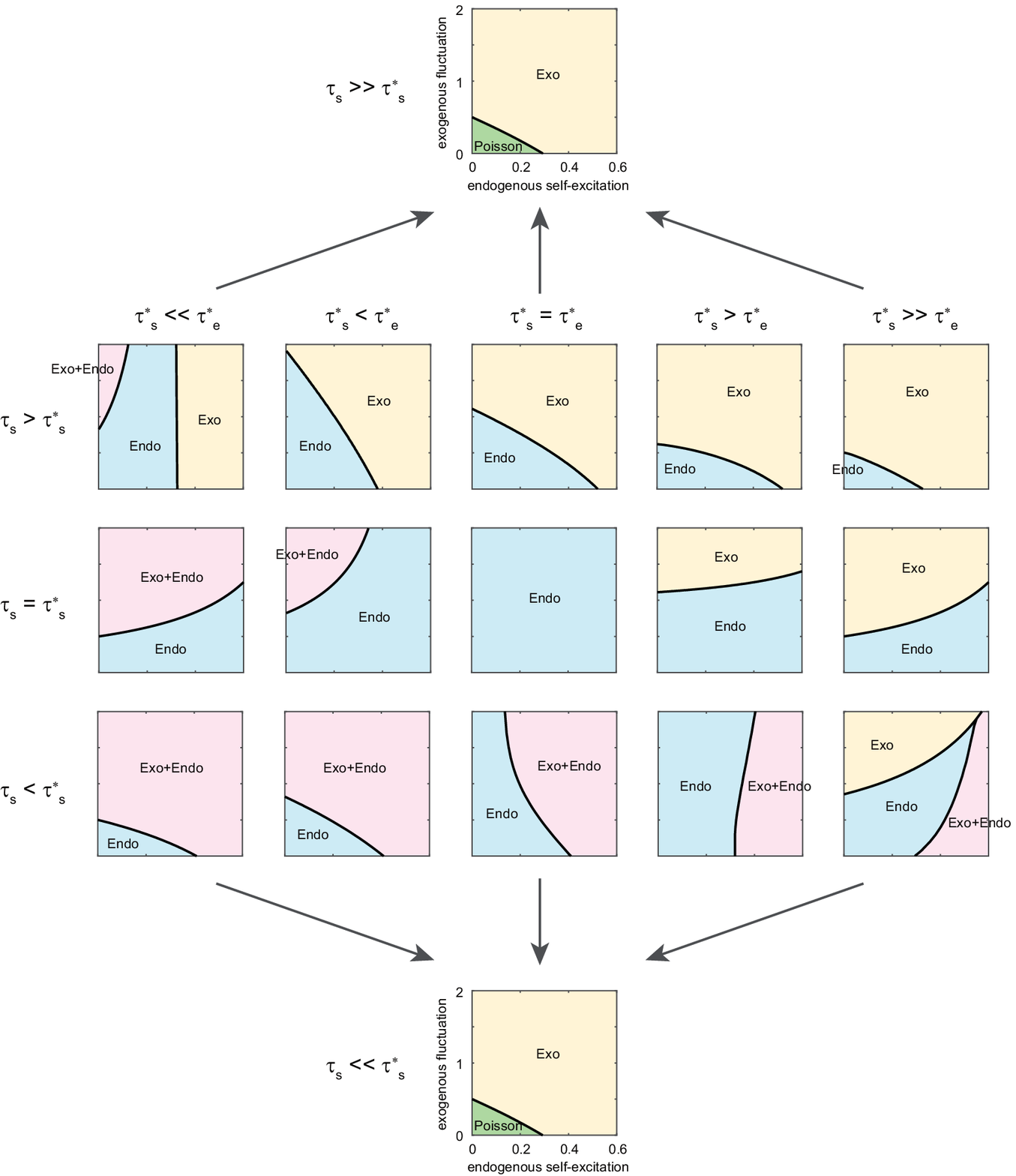}
\caption{
Transformation of the phase diagram of the Hawkes decoder, according to the change in the timescales of the decoder's kernel, $\tau_s$, the original self-excitation kernel, $\tau^*_s$, and the exogenous fluctuation, $\tau^*_e$. 
}
\label{fig:phasediagram_entire}
\end{center}
\end{figure*}
%%%%%%%%%%%%%%%%%%%%%%%%%%%%%%%%%%%%%%%%%%%%%%%%%%%%%%%%%%

%%%%%%%%%%%%%%%%%%%%%%%%%%%%%%%%%%%%
\subsubsection*{The Hawkes decoder with a correct self-excitation kernel}

First, we analyze the synthetic data using the Hawkes decoder. We consider here the case in which the model uses the correct shape of the original excitation kernel, $\phi (t) = \phi^*(t)$. The strength of self-excitation $\alpha$ and the hyperparameter $\gamma$ for estimating the external fluctuation are selected by minimizing the free energy, $F(\alpha,\gamma)$, Eq.~(\ref{eq:free_energy_def}). 

Figure~\ref{fig:phasediagram_hawkes_poisson} (a) depicts a phase diagram of the Hawkes decoder in a plane spanned by the parameters of an event-generation model, the reproduction ratio of the endogenous self-excitation $\alpha^*$, and the strength of exogenous fluctuation ${\sigma^*}^2 \tau_e^*/\mu^*$, for the case of $\tau_e^* \gg \tau^*_s$. We obtained three regimes: $\{\hat{\alpha}=0, \hat{\gamma} =0\}$ (``Poisson''), $\{\hat{\alpha}\ne 0, \hat{\gamma} =0\}$ (``Endo''), and $\{\hat{\alpha}\ne 0, \hat{\gamma} \ne 0\}$ (``Exo+Endo''). In Fig.~\ref{fig:phasediagram_hawkes_poisson} (b), we show sample event series derived from the inhomogeneous Hawkes process and the rates $\hat{\lambda}(t)$ estimated via the Empirical Bayes method.

The $y$-axis ($\alpha^* = 0$) in Fig.~\ref{fig:phasediagram_hawkes_poisson} (a) represents the situation by which events are generated by the inhomogeneous Poisson process in which the rate $\nu^* (t)$ fluctuates with a standard deviation $\sigma^*$ around the mean rate $\mu^*$ in a timescale of $\tau_e^*$. Because the event-generation process is stochastic in nature, the profile of the underlying rate $\nu^* (t)$ cannot be captured precisely from a series of occurrence times, and it might even occur that the model abandons to capture a fluctuation if the amplitude of the original rate fluctuation is too small. This $\hat{\gamma} = 0$ occurs if estimating a fluctuating rate (with $\gamma \ne 0$) is likely to produces a larger error than indicating a constant rate (with $\gamma = 0$). We found that the model suggests a constant rate if the original rate fluctuation is small (Appendix \ref{appendix:correct_self-excitation_kernel}):
\begin{equation}
{\sigma^*}^2 \tau_e^*/\mu^* < \frac{1}{2(1-\alpha^*)}.
\label{eq:gamma0} 
\end{equation}

We have considered minimizing the mean integrated squared error between the underlying rate and the histogram of a given series of event times and discovered that the optimal bin size may diverge when the underlying fluctuation is small~\cite{Koyama2004}. For the doubly stochastic Poisson process Eq.~(\ref{eq:oup}), the condition in which the bin size diverges is identical to inequality (\ref{eq:gamma0}) with $\alpha^* = 0$. This implies that the condition for the rate fluctuation being unknowable is independent of the estimation methods.

%%%%%%%%%%%%%%%%%%%%%%%%%%%%%%%%%%%%
\subsubsection*{The Bayesian rate estimator}

Now we analyze the identical dataset using the ``Bayesian rate estimator,'' which estimates the fluctuating rate $\nu(t)$ by fitting the inhomogeneous Poisson process to the data. In this case, we compute the marginal likelihood by setting $\alpha=0$ in Eq.~(\ref{eq:marginal_likelihood}), and obtain the free energy as
\begin{eqnarray}
F_{\rm p}(\gamma) = F(\alpha=0,\gamma). 
\label{eq:free_energy_poisson}
\end{eqnarray}
In this case, the process of estimating $\{\nu(t)\}$ is identical to the rate estimation method that we have developed previously~\cite{Koyama2007}. 

Figure~\ref{fig:phasediagram_hawkes_poisson} (c) depicts a phase diagram of the Bayesian rate estimator for the same data examined with the Hawkes decoder in Fig.~\ref{fig:phasediagram_hawkes_poisson} (a). Here we obtain two different regimes categorized according to whether $\hat{\gamma} =0$ (``Poisson'') or $\hat{\gamma} \ne 0$ (``Exo'').

The $x$-axis of the diagram (${\sigma^*}^2 \tau_e^*/\mu^*=0$) in Fig.~\ref{fig:phasediagram_hawkes_poisson} (c) represents the data generated by the (homogeneous) Hawkes process. Even if the exogenous fluctuation is absent, an apparent occurrence of events may exhibit the large fluctuation due to the self-excitation mechanism. We found that the Bayesian rate estimator may suggest a fluctuating rate $\hat{\nu}(t)$ or $\hat{\gamma} \ne 0$ if the original self-excitation is greater than the critical value (Appendix \ref{appendix:bayesian_rate_estimator}):
\begin{equation}
\alpha^* > \alpha_c = 1-1/\sqrt{2} \approx 0.2929.
\end{equation}
The critical reproduction ratio $\alpha_c$ is identical to that obtained for the principled histogram method, such that the selected bin size diverges above this reproduction ratio $\alpha_c$~\cite{Onaga2014, Onaga2016}.

As shown in Fig.~\ref{fig:phasediagram_hawkes_poisson} (c), the Bayesian rate estimator cannot capture the rate fluctuation if the endogenous self-excitation $\alpha^*$ or exogenous fluctuation ${\sigma^*}^2 \tau_e^*/\mu^*$ is small. In Fig.~\ref{fig:phasediagram_hawkes_poisson} (d), we also show the rates $\hat{\nu}(t)$ estimated by the Bayesian rate estimator for the same series of events analyzed by the Hawkes decoder in Fig.~\ref{fig:phasediagram_hawkes_poisson} (b).

%%%%%%%%%%%%%%%%%%%%%%%%%%%%%%%%%%%%
\subsubsection*{The Hawkes decoder with an incorrect self-excitation kernel}

Even if event occurrence is facilitated by the self-excitation mechanism, the self-excitation information is generally not available. We consider the case in which the Hawkes decoder assumes a kernel that is different from that of the original process. Here in particular, we analyze the case in which the timescale $\tau_s$ of the self-exciting kernel $\phi(t) = 1/\tau_s \exp(-t/\tau_s)$ is different from the timescale $\tau^*_s$ of the original kernel $\phi^*(t) = 1/\tau^*_s \exp(-t/\tau^*_s)$, and see how the phase diagram is modified by choosing an incorrect self-excitation kernel (here we examine the case of $\tau_e^* \gg \tau^*_s$).

Figure~\ref{fig:phasediagram_hawkes_wrong} (c) represents the phase diagram obtained with a correct kernel, $\tau_s = \tau^*_s$. If the timescale of the decoder's kernel $\tau_s$ is longer than the original, $\tau_s > \tau^*_s$, the ``Exo+Endo'' regime reduces and the ``Exo" regime emerges from the right-hand side of the phase diagram (Fig.~\ref{fig:phasediagram_hawkes_wrong} (b)). If the timescale of the decoder's kernel is even longer, the ``Exo" regime dominates, and the self-excitation is not detected. In the limit of $\tau_s \gg \tau^*_s$, its phase diagram is identical to that of the Bayesian rate estimator (Fig.~\ref{fig:phasediagram_hawkes_wrong} (a)). 

In the case where the timescale of the decoder's kernel is shorter than the original, $\tau_s < \tau^*_s$, the Hawkes regime expands (Fig.~\ref{fig:phasediagram_hawkes_wrong} (d)) from the regime obtained with a correct kernel (Fig.~\ref{fig:phasediagram_hawkes_wrong} (c)). In this case, the reproduction ratio $\alpha$ is estimated to be smaller than the original ($\hat{\alpha} < \alpha^*$). In the limit of $\tau_s \ll \tau^*_s$, the estimated reproduction ratio becomes infinitesimal, and accordingly the ``Endo'' and ``Exo+Endo'' regimes turn into ``Poisson'' and ``Exo'' regimes, respectively (Appendix \ref{appendix:incorrect_self-excitation_kernel}). In this limit, the phase diagram is also identical to that of the Bayesian rate estimator (Fig.~\ref{fig:phasediagram_hawkes_wrong} (e)).

%%%%%%%%%%%%%%%%%%%%%%%%%%%%%%%%%%%%
\subsubsection*{The case of slow self-excitation}

So far we have examined the cases in which an environmental factor changes slowly compared to the self-excitation ($\tau_e^* \gg\tau^*_s$), and shown that the Hawkes decoder can estimate the contributions of exogenous and endogenous factors (Fig.~\ref{fig:phasediagram_hawkes_timescale} (a)). If the timescale of exogenous fluctuation $\tau_e^*$ is comparable or even shorter than that of self-excitation $\tau^*_s$, however, it is difficult to separate the exogenous and endogenous contributions to the data.

If the timescale of the external fluctuation $\tau_e^*$ is relatively close to that of the self-exciation $\tau^*_s$, the ``Exo+Endo'' regime decreases (Fig.~\ref{fig:phasediagram_hawkes_timescale} (b)). If the timescale of the extrinsic fluctuation is comparable to that of self-excitation $\tau_e^* = \tau^*_s$, we may not discriminate between the exogenous and endogenous self-excitation components and obtain only the ``Endo'' regime (Fig.~\ref{fig:phasediagram_hawkes_timescale} (c)). Contrary, if the timescale of the external fluctuation $\tau_e^*$ is shorter than that of the self-exciation $\tau^*_s$, the decoder may yield the ``Exo'' regime (Figs.~\ref{fig:phasediagram_hawkes_timescale} (d) and (e)). 

Figure \ref{fig:phasediagram_entire} depicts the full view of the transition of the phase diagram, consisting of a variety of situations.

%%%%%%%%%%%%%%%%%%%%%%%%%%%%%%%%%%%%
\subsection{Analyzing real-world data}
%%%%%%%%%%%%%%%%%%%%%%%%%%%%%%%%%%%%
%%%%%%%%%%%%%%%%%%%%%%%%%%%%%%%%%%%%

%%%%%%%%%%%%%%%%%%%%%%%%%%%%%%%%%%%%%%%%%%%%%%%%%%%%%%%%%%
\begin{figure*}[htbp]
\begin{center}
\includegraphics[width=17.2cm]{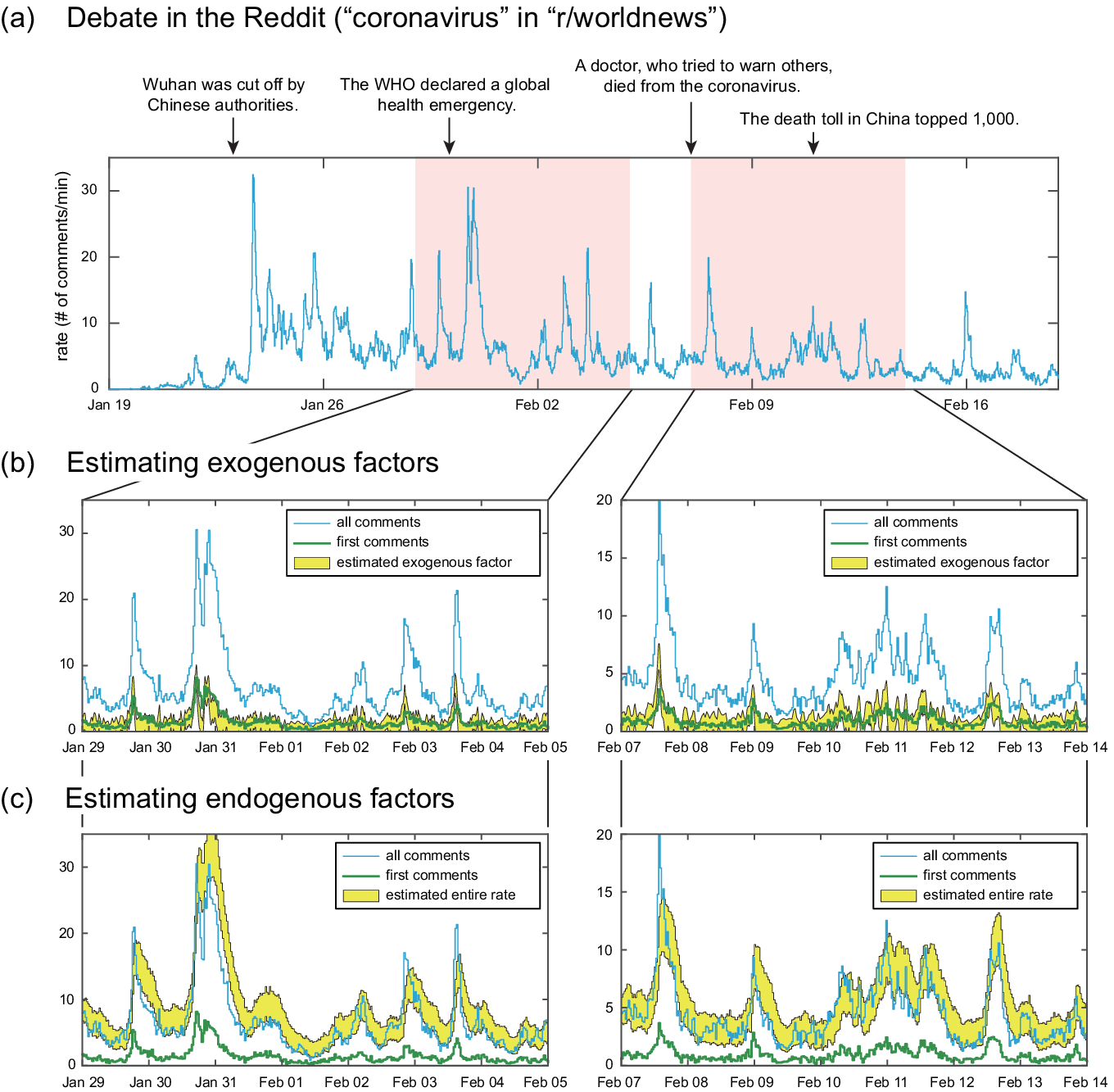}
\caption{
Analysis of real-world data. (a) The number of comments containing the keyword ``coronavirus'' submitted to the Reddit forum ``r/worldnews'' for a period between January 19th and February 19th, 2020. (b) Estimating the exogenous factor representing environmental changes from the entire event series. Submission activities for two time series of 1 week (between Jan 29 and Feb 5 and between Feb 7 and Feb 14). The blue and red lines indicate the rate of all submissions and the first comments followed by the original posts, respectively. The yellow shaded areas represent the 95\% range of exogenous contribution estimated by the Hawkes decoder. (c) Estimating the endogenous self-excitation causing chains of events, given the environmental changes. The entire occurrence rate $\lambda(t)$ was obtained by simulating the inhomogeneous Hawkes process, by considering the occurrence rate of first comments as the exogenous input $\nu(t)$. The reproduction ratio $\alpha$ and the timescale of the self-excitation $\tau_s$ are obtained by fitting the Hawkes decoder for the previous dataset. 
The yellow shaded areas represent the 95\% range of estimated entire occurrence rate.
}
\label{fig:result_reddit}
\end{center}
\end{figure*}
%%%%%%%%%%%%%%%%%%%%%%%%%%%%%%%%%%%%%%%%%%%%%%%%%%%%%%%%%%

\subsubsection*{Time series of comments submitted to an SNS}

Finally, we analyze real-world data, namely the time series of comments talking about a given subject on a particular Reddit forum, either in original posts or in comments upon them. Data were collected through the public API for the keyword ``coronavirus'' in the subreddit ``r/worldnews'' for a period between January 19th and February 19th, 2020. The dataset contains 223,545 comments in response to 5,341 submissions. Figure~\ref{fig:result_reddit} (a) displays the changes in commenting activity over a month, indicating a strong burst of comments occurring at many times, presumably induced by the news.

When considering topic-related content, the first comments on original posts might be interpreted as a direct consequence of exogenous influences, because they were likely to be induced by real-world events. The other follow-up comments may be considered as endogenous activity that was induced within a forum. 

We applied the Hawkes decoder to the time series of times of all the comments (whether first comments or follow-up comments) to estimate the exogenous and endogenous factors that have worked to generate events. Due to the large size of the observation window, we selected several 1-week datasets from the full time series. When applying to the Empirical Bayes method, we also selected the timescale of the self-exciting kernel so that the likelihood is maximized. By analyzing several 1-week time series, the timescale of the kernel was selected in a range between 1,300 and 3,000 sec, implying that commenting may typically be done in about half an hour. This is in contrast to our previous GLM analysis of the tweets that contain a hashtag related to ``bitcoin,'' in which case we used a kernel of the timescale $60$ seconds or 1 min. This short response-time is presumably because clicking the retweet button can be done quickly, in contrast to the submission in Reddit, for which one needs some time to organize a comment. 

The results of the decoding analysis for two segments are shown in Fig.~\ref{fig:result_reddit} (b). The rates of total submissions and the original posts and direct comments were shown using 30-min binning. The exogenous rate estimated by the Hawkes decoder was superimposed upon the figure. Here we plotted the 95\% range of $\nu(t)$ estimated from the posterior distribution $p_{\hat{\alpha},\hat{\gamma}}(\{\nu(t)\}|\{t_i\})$ computed by Eq.~(\ref{eq:posterior}) with the parameter $\hat{\alpha}$ and hyperparameter $\hat{\gamma}$ determined by Eq.~(\ref{eq:estimate_hyperparameters}). We observe that the estimated exogenous component $\nu(t)$ exhibits good agreement with the rate of first comments. The reproduction ratio $\hat{\alpha}$ was estimated as $0.84$ and $ 0.81$, and the timescale $\tau$ was selected as 1,644 and 2,456 sec for these two segments.

%%%%%%%%%%%%%%%%%%%%%%%%%%%%%%%%%%%%
\subsubsection*{Predicting chain-reactions induced by external influence}

While the Hawkes decoder can discover exogenous and endogenous factors in a single event series, its validity cannot be proven rigorously as long as the information about these origins is unavailable. However, it may be possible to use a model for predicting chain-reactions that are indirectly induced by environmental changes.

We analyzed the commenting data in the Reddit. By considering the occurrence rate of first comments as the external influence $\nu(t)$, we simulated the Hawkes process to estimate the entire occurrence rate $\lambda(t)$. To do this, we estimated the reproduction ratio $\alpha$ and the timescale of the self-excitation kernel $\tau_s$ by fitting the Hawkes decoder to the data of the preceding one week. With the series of events generated by this procedure, we constructed a time histogram of the occurrence rate with a bin size of 30 min. By repeating this procedure 1,000 times for the same $\nu(t)$, we obtained 1,000 different time histograms, with which we can obtain the distribution of the number of comments. 

Figure \ref{fig:result_reddit} (c) depicts two examples of commenting activity for 1 week, with the distribution of occurrence rates $\lambda(t)$ predicted from the first comments. The rate of all comments that have occurred in practice is plotted on the distribution, exhibiting good predictive performance.

%%%%%%%%%%%%%%%%%%%%%%%%%%%%%%%%%%%%
\section{Discussion}
%%%%%%%%%%%%%%%%%%%%%%%%%%%%%%%%%%%%

The Hawkes process has attracted attention as a mathematical model that may describe the self-excitation mechanisms generating a chain of events, and there have been many attempts to fit the model to given event times. 

Firstly, the time dependency of self-excitation has been pursued by fitting the original (homogeneous) Hawkes process to a given series of event-occurrence times. The estimation was performed by maximizing the likelihood using various analytical techniques, either parametrically~\cite{Ogata1978, Ozaki1979, Veen2008, Rasmussen2013, Fonseca2014, Chen2018} or nonparametrically~\cite{Lewis2011, Bacry2012, Bacry2016a, Bacry2016b, Patricia2010, Hansen2015, zhou13, xuc16, Kirchner2016, Kirchner2017} (an exhaustive review is given in~\cite{bacry2015}). 

Secondly, potential environmental changes were taken into account by introducing temporal fluctuation to the background rate. There are several attempts to estimate the background rate nonparametrically, including an Expectation-Maximization (EM) algorithm~\cite{Lewis2011}, kernel-density estimation and differential-entropy minimization~\cite{chen_hall_2016}, and local-maximum-likelihood estimation~\cite{Clinet2018}. There have also been methods for estimating the background rate parametrically~\cite{chen_hall_2013, kobayashi2016, Omi2017}.

Our analytical method can be categorized into the latter group of studies; we have developed a method of estimating both exogenous and endogenous factors by fitting a linear superposition of the inhomogeneous Poisson process and the Hawkes process to an observed sequence of event times. With synthetic data generated by the inhomogeneous Hawkes process, we have confirmed that the method works properly for estimating the original parameters. 

Here in particular, we have found that there are cases in which even the best decoding method cannot capture the extrinsic fluctuation and/or intrinsic self-excitation. Regarding the extrinsic fluctuation, a principled estimator may assume a constant environment if the extrinsic fluctuation is too small: even though the decoding method can estimate the fluctuating rate from a given dataset, the estimation error may become larger than that obtained by assuming a constant rate. Regarding the intrinsic self-excitation, the model cannot separate the self-excitation from environmental fluctuation if the timescale of excitation is similar to or larger than that of environmental fluctuation. We have summarized the undetectability conditions up into phase diagrams categorizing four regimes according to whether or not the exogenous fluctuation and endogenous self-excitation are respectively detected.

We also devised the Hawkes decoder to estimate the exogenous and endogenous factors from a given series of events. By applying it to real time series of debate on Reddit, we have observed that the first comments and the follow-up comments map closely to the estimated exogenous and endogenous reactions, respectively.

While the Hawkes decoder can estimate the contributions of exogenous and endogenous factors in a single event series, it is often the case that information about the origin is unavailable. In such a case, the action of dividing the underlying causes into exogenous and endogenous categories might be regarded as a subjective interpretation. However, the estimation of respective contributions might be useful if it correctly predicts the impact of external factors upon the resulting occurrence. For the real time series of debate on Reddit, we considered the occurrence rate of first comments as the exogenous influence, and simulated the Hawkes process to ``predict'' the number of follow-up comments. We confirmed that the prediction exhibited a good agreement with the number of follow-up comments that occur in practice.

Our method is general and applicable to a variety of data. We have provided an application program, with which anyone can analyze any series of event times.

%%% Appendix %%%%%%%%%%%%%%%%%%%%%%%%%%%%%%%%%%%%%%%%%%%%%%%%%%%%%%%%%%%%%%%%%%
%%%%%%%%%%%%%%%%%%%%%%%%%%%%%%%%%%%%%%%%%%%%%%%%%%%%%%%%%%%%%%%%%%%%%%%%%%%%%
%%%%%%%%%%%%%%%%%%%%%%%%%%%%%%%%%%%
\appendix
\section{Numerical method}
\label{appendix:numerical_method}
%%%%%%%%%%%%%%%%%%%%%%%%%%%%%%%%%%%

We reformulate the Hawkes decoder model, defined by Eqs.~(\ref{eq:pdf_hawkes}) and (\ref{eq:prior}), as a state-space model, based on which a numerical method for selecting the parameter and hyperparameter $\{\hat{\alpha},\hat{\gamma}\}$ and a time-dependent exogenous stimulus $\hat{\nu}(t)$ are developed. 

%%%%%%%%%%%%%%%%%%%%%%%%
\subsection{State-space representation}

We use the exponential function $\phi(t) = (1/\tau_s)\exp(-t/\tau_s)$ for the self-excitation kernel.
From Eqs.~(\ref{eq:hawkes}) and (\ref{eq:discrete}), the probability density of 
the inter-event interval $t_i-t_{i-t}$ is expressed as 
\begin{eqnarray}
\lefteqn{
\left( \nu^+(t_i) + \frac{\alpha}{\tau_s}R_i \right)
}\nonumber\\
&& \times
\exp\left( -\int_{t_{i-1}}^{t_i}\nu^+(t)dt - \alpha R_i(e^{(t_i-t_{i-1})/\tau_s}-1) \right),
\label{eq:conditional_pdf}
\end{eqnarray}
where $\nu^+(t) = \max(0,\nu(t))$, which ensures non-negativity of the exogenous stimulus, and $R_i$ is computed for $i=2,\ldots,n$ as 
\begin{equation}
R_i = (1+R_{i-1})e^{-(t_i-t_{i-1})/\tau_s}, 
\label{eq:R_i}
\end{equation}
with an initial value $R_1$. We approximate the exogenous stimulus as being piecewise constant, 
\begin{equation}
\nu(t) \approx \nu_i, \quad t_{i-1} < t \le t_i,
\end{equation}
which is valid under the condition that the timescale of the exogenous fluctuation is sufficiently large enough compared with the mean inter-event interval. 
Under this approximation, and letting $y_i=t_i-t_{i-1}$ be the $i$th inter-event interval, 
Eq.~(\ref{eq:conditional_pdf}) becomes
\begin{eqnarray}
p_{\alpha}(y_i|\nu_i) 
&=& \left( \nu_i^+ + \frac{\alpha}{\tau_s}R_i \right) \nonumber\\
&&\times
\exp(
-y_i\nu_i^+ - \alpha R_i(e^{y_i/\tau_s}-1)
),
\label{eq:observation}
\end{eqnarray}
which we consider the conditional density of $y_i$ given $\nu_i$.

The transition density of $\nu_i$ associated with the prior distribution (\ref{eq:prior}) is given by
\begin{equation}
p_{\gamma}(\nu_i|\nu_{i-1}) = 
\frac{1}{\sqrt{\pi\gamma^2(t_i-t_{i-2})}}\exp\left(-\frac{(\nu_i-\nu_{i-1})^2}{\gamma^2(t_i-t_{i-2})}\right).
\label{eq:state}
\end{equation}
Combined with an initial density $p(\nu_1)$, Eqs.~(\ref{eq:observation}) and (\ref{eq:state}) define a state-space model~\cite{West1997,Durbin2001}, 
for which the empirical Bayes method can be implemented by the recursive Bayesian algorithm.

%%%%%%%%%%%%%%%%%%%%%%%%%%%%%%%%%%%
\subsection{Recursive Bayesian algorithm}

For notational simplicity, let $Y_i = \{y_1,\ldots, y_i\}$ be the observations up to time $t_i$.
By the Bayes' theorem, the posterior distribution of $\nu_i$, given the observations up to the current time, is expressed as
\begin{equation}
p_{\alpha,\gamma}(\nu_i|Y_i) = 
\frac{p_{\alpha}(y_i|\nu_i) p_{\alpha,\gamma}(\nu_i|Y_{i-1})}{\int p_{\alpha}(y_i|\nu_i) p_{\alpha,\gamma}(\nu_i|Y_{i-1})d\nu_i}, 
\label{eq:filtering}
\end{equation}
where $p_{\alpha,\gamma}(\nu_i|Y_{i-1})$ on the right-hand-side
is computed using the posterior distribution from the last iteration, $p_{\alpha,\gamma}(\nu_{i-1}|Y_{i-1})$, as 
\begin{equation}
p_{\alpha,\gamma}(\nu_i|Y_{i-1}) = \int p_{\gamma}(\nu_i|\nu_{i-1})p_{\alpha,\gamma}(\nu_{i-1}|Y_{i-1})d\nu_{i-1}.
\label{eq:predictive}
\end{equation}
Thus, starting with the initial distribution $p_{\alpha,\gamma}(\nu_1|Y_0)= p(\nu_1)$, the posterior distributions (\ref{eq:filtering}) and (\ref{eq:predictive}) are recursively computed for $i=1,\ldots, n$. 
Once we obtain these distributions, the posterior distribution of $\nu_i$, given the whole observation $Y_n$, is computed using the following recursive equation, 
\begin{eqnarray}
p_{\alpha,\gamma}(\nu_i|Y_n) 
&=& 
\int \frac{p_{\alpha,\gamma}(\nu_{i+1}|Y_n) p_{\gamma}(\nu_{i+1}|\nu_i)}{p_{\alpha,\gamma}(\nu_{i+1}|Y_i)}d\nu_{i+1} \nonumber\\
&&\times
p_{\alpha,\gamma}(\nu_i|Y_i),
\label{eq:smoothing}
\end{eqnarray}
for $i=n-1,\ldots,1$ in backward. 
We obtain the MAP estimate of the exogenous stimulus $\{\hat{\nu}_i\}$, such that $p_{\alpha,\gamma}(\nu_i|Y_n)$ (for $i=1,\ldots,n$) is maximized. 

For the state-space model (\ref{eq:observation}) and (\ref{eq:state}), 
we introduce a Gaussian approximation in the posterior distribution (\ref{eq:filtering}) at each point in time, providing a simple algorithm that is computationally tractable \cite{Koyama2010JASA,Koyama2010JCNS,Koyama2010AISM}.
Let $\nu_{i-1|i-1}$ and $q_{i-1|i-1}$ be the (approximate) mean and variance for $p_{\alpha,\gamma}(\nu_{i-1}|Y_{i-1})$. 
Under a Gaussian approximation in $p_{\alpha,\gamma}(\nu_{i-1}|Y_{i-1})$, the posterior distribution (\ref{eq:predictive}) is also a Gaussian whose mean and variance are, respectively, computed as 
\begin{eqnarray}
\nu_{i|i-1} &=& \nu_{i-1|i-1}, \label{eq:predictive_mean} \\
q_{i|i-1} &=& q_{i-1|i-1} + \gamma^2(t_i-t_{i-2})/2. \label{eq:predictive_var}
\end{eqnarray}

To make a Gaussian approximation in Eq.~(\ref{eq:filtering}), let $l(\nu_i) = \log p_{\alpha}(T_i|\nu_i) p_{\alpha,\gamma}(\nu_i|T_{1:i-1})$ be the log posterior distribution (from which we omit the normalization constant). 
Expanding the log posterior distribution about its maximum point up to the second-order yields 
$l(\nu_i) \approx l(\nu_{i|i}) + \ddot{l}(\nu_{i|i})(\nu_i - \nu_{i|i})^2/2$, 
where $\nu_{i|i}$ is obtained as a root of the equation $\dot{l}(\nu_i)=0$,
\begin{eqnarray}
\nu_{i|i} &=& 
\big(
\nu_{i|i-1} - \alpha/\tau_sR_i - y_iq_{i|i-1} \nonumber\\
&& +
\{
(\nu_{i|i-1} - \alpha/\tau_sR_i - y_iq_{i|i-1})^2 \nonumber\\
&& +4(q_{i|i-1} + \alpha/\tau_sR_i\nu_{i|i-1} \nonumber\\
&& -\alpha/\tau_sR_iy_iq_{i|i-1})
\}^{1/2}
\big) / 2.
\label{eq:filtering_mean}
\end{eqnarray}
Thus the posterior distribution (\ref{eq:filtering}) is approximated to a Gaussian with mean $\nu_{i|i}$ and variance:
\begin{eqnarray}
q_{i|i} &=& -\ddot{l}(\nu_{i|i})^{-1} \nonumber\\
&=&
\frac{q_{i|i-1}(\nu_{i|i}+\alpha/\tau_sR_i)^2}{q_{i|i-1}+(\nu_{i|i}+\alpha/\tau_sR_i)^2}.
\label{eq:filtering_var}
\end{eqnarray} 

The Gaussian approximations in $p_{\alpha,\gamma}(\nu_i|Y_i)$ and $p_{\alpha,\gamma}(\nu_i|Y_{i-1})$ result in a Gaussian for Eq.~(\ref{eq:smoothing}) as well. 
Let $\nu_{i|n}$ and $q_{i|n}$ be the mean and variance of $p_{\alpha,\gamma}(\nu_i|Y_n)$. We then obtain the recursive equation corresponding to Eq.~(\ref{eq:smoothing}):
\begin{eqnarray}
\nu_{i|n} &=& \nu_{i|i} + (\nu_{i+1|n}-\nu_{i+1|i}) q_{i|i}/q_{i+1|i}, \label{eq:smoothing_mean}\\
q_{i|n} &=& q_{i|i} + (q_{i+1|n}-q_{i+1|i}) (q_{i|i}/q_{i+1|i})^2. \label{eq:smoothing_var}
\end{eqnarray}

Eqs.~(\ref{eq:predictive_mean})--(\ref{eq:smoothing_var}) comprise the recursive estimation of $\{\nu_i\}$. 
The algorithm is summarized as follows. 
\begin{enumerate}
\item 
Set initial values $\nu_{1|0}$, $q_{1|0}$ and $R_1$.
\item \label{item:step_2}  
Compute Eqs.~(\ref{eq:R_i}) and (\ref{eq:predictive_mean})--(\ref{eq:filtering_var}) for $i=1,\ldots,n$ in forward.
\item 
Starting with $\nu_{n|n}$ and $q_{n|n}$,
which are obtained at the last iteration in the step \ref{item:step_2}, 
compute Eqs.~(\ref{eq:smoothing_mean}) and (\ref{eq:smoothing_var}) for $i=n-1,\ldots,1$ in backward.
\end{enumerate}
The resulting $\{\nu_{i|n}\}$ provides the MAP estimate of the exogenous rate.
Note that we may use a diffuse (noninformative) prior for the initial values ($q_{1|0}\to\infty$), which results in $\nu_{1|1}=1/y_1-\alpha/\tau_sR_1$ and $q_{1|1}=1/y_1^2$, leaving out the dependency of the initial values upon the estimation. In our analysis, we used $R_1=\tau_s/\overline{y}$, where $\overline{y}=\sum_{i=1}^my_i/m$ is an average of the observations over a given range (we have chosen $m=100$), to remove the initial non-stationary part of the estimation. 

To select the parameter and hyperparameter $\{\hat{\alpha},\hat{\gamma}\}$, we consider a factorization of the marginal likelihood, 
\begin{eqnarray}
p_{\alpha,\gamma}(Y_n) &=&
\prod_{i=1}^np_{\alpha,\gamma}(y_i|Y_{i-1}) \nonumber\\
&=&
\prod_{i=1}^n \int_{-\infty}^{\infty} d\nu_i p_{\alpha}(y_i|\nu_i) p_{\alpha,\gamma}(\nu_i|Y_{i-1}).
\label{eq:marginal_likelihood_factorized}
\end{eqnarray}
Since $p_{\alpha,\gamma}(\nu_i|Y_{i-1})$ on the right-hand-side is approximated by a Gaussian with mean $\nu_{i|i-1}$ and variance $q_{i|i-1}$, the integral may be approximated by the Gauss-Hermite quadrature,
\begin{eqnarray}
\lefteqn{\int_{-\infty}^{\infty} d\nu_i p_{\alpha}(y_i|\nu_i) p_{\alpha,\gamma}(\nu_i|Y_{i-1})} \hspace{0.1cm} \nonumber\\
&=&
\frac{1}{\sqrt{2\pi q_{i|i-1}}}
\int_{-\infty}^{\infty} d\nu_i p_{\alpha}(y_i|\nu_i) 
\exp\left( -\frac{(\nu_i-\nu_{i|i-1})^2}{2q_{i|i-1}} \right) \nonumber\\
&=&
\frac{1}{\sqrt{\pi}} \int_{-\infty}^{\infty} dx
p_{\alpha}(y_i|\sqrt{2q_{i|i-1}}x+\nu_{i|i-1}) e^{-x^2}
\nonumber\\
&\approx&
\frac{1}{\sqrt{\pi}}\sum_{l=1}^m w_l p_{\alpha}(y_i|\sqrt{2q_{i|i-1}}x_l+\nu_{i|i-1}) ,
\end{eqnarray}
where the weights $\{w_l\}$ and evaluation points $\{x_l\}$ are chosen according to a quadrature rule \cite{Press1992}.
The parameter and hyperparameter $\{\hat{\alpha},\hat{\gamma}\}$ are selected by maximizing Eq.~(\ref{eq:marginal_likelihood_factorized}) numerically. 
It should be noted that the time constant of the self-excitation kernel, $\hat{\tau}_s$, may also be selected by maximizing Eq.~(\ref{eq:marginal_likelihood_factorized}) with respect to $\tau_s$, as well.

%%%%%%%%%%%%%%%%%%%%%%%%%%%%%%%%%%%
\section{Derivation of the free energy}
\label{appendix:derivation_free_energy}
%%%%%%%%%%%%%%%%%%%%%%%%%%%%%%%%%%%
\subsection{Representation of intensity}

First, we represent the intensity (\ref{eq:hawkes_original}) of the inhomogeneous Hawkes process in terms of the mean behavior and the fluctuations. 
We consider decomposing a series of events into the mean and fluctuation as
\begin{equation}
\sum_i\delta(t-t_i) = \lambda^*(t) + \xi(t),
\label{eq:martingale_decomposition}
\end{equation}
where $\xi(t)$ is the white noise (the ``derivative" of the martingale \cite{Bacry2012}), whose ensemble characteristics satisfy $\langle \xi(t) \rangle = 0$ and $\langle \xi(t)\xi(t') \rangle = \langle \lambda^*(t) \rangle \delta(t-t')$.
Using this decomposition, Eq.~(\ref{eq:hawkes_original}) can be represented as
\begin{eqnarray}
\lambda^*(t) 
&=&
\nu^*(t) + \alpha^*\int_0^t\phi^*(t-u)\lambda^*(u)du \nonumber\\
&&+ \alpha^*\int_0^t\phi^*(t-u)\xi(u)du.
\end{eqnarray}
By applying the Laplace transformation, this equation is solved as
\begin{eqnarray}
\lambda^*(t) &=& \nu^*(t) + \int_0^t\Psi^*(t-u)\nu^*(u)du \nonumber\\
&&+ \int_0^t\Psi^*(t-u)\xi(u)du,
\label{eq:conditional_intensity_true_new}
\end{eqnarray}
where
$\Psi^*(t)$ is an ``effective self-exciting kernel" whose Laplace transform is given by 
\begin{equation}
\widehat{\Psi}^*(s) = \frac{\alpha^*\widehat{\phi}^*(s)}{1-\alpha^*\widehat{\phi}^*(s)},
\label{eq:effective_kernel_true}
\end{equation}
where $\widehat{\phi}^*(s)$ is the Laplace transform of the self-excitation kernel $\phi^*(t)$. 
The first and second terms on the right-hand-side of Eq.~(\ref{eq:conditional_intensity_true_new}) represent the average behavior and the third term represents the fluctuation around the averaged behavior.
Representing the original exogenous input (\ref{eq:oup}) as 
$\nu^*(t) = \mu^* + \sigma^*\zeta(t)$, 
where $\zeta(t)$ is the normalized fluctuation whose autocorrelation function is given by $\langle \zeta(t)\zeta(t+u) \rangle = \exp(-|u|/\tau^*_e)$, 
Eq.~(\ref{eq:conditional_intensity_true_new}) is expressed as
\begin{equation}
\lambda^*(t) = \Lambda^* + \sigma^*\zeta(t) + \sigma^* Z^*(t) + \Xi^*(t),
\label{eq:conditional_intensity_true_decomposition}
\end{equation}
where
\begin{eqnarray}
\Lambda^* &=& \langle \lambda^*(t) \rangle = \frac{\mu^*}{1-\alpha^*}, \label{eq:mean} \\
Z^*(t) &=& \int_0^t\Psi^*(t-u)\zeta(u)du, \\
\Xi^*(t) &=& \int_0^t\Psi^*(t-u)\xi(u)du.
\end{eqnarray}

In the same manner, by decomposing the exogenous rate in the decoder into the mean and fluctuation, $\nu(t)= (1-\alpha)\Lambda^*+x(t)$, 
the decoder's intensity (\ref{eq:hawkes}) is represented as
\begin{equation}
\lambda(t) = \Lambda^* + x(t)+ \sigma^*Z(t) + \Xi(t),
\label{eq:conditional_intensity_decoder_decomposition}
\end{equation}
where
\begin{eqnarray}
Z(t) &=& \int_0^t \Psi(t-u)\zeta(u)du, \\
\Xi(t) &=& \int_0^t\Psi(t-u)\xi(u)du.
\end{eqnarray}
Here, $\Psi(t)$ represents the effective self-exciting kernel of the decoder, whose Laplace transform is given by 
\begin{eqnarray}
\widehat{\Psi}(s) = \frac{\alpha \widehat{\phi}(s)}{1-\alpha^*\widehat{\phi}^*(s)},
\label{eq:effective_kernel_model}
\end{eqnarray}
where $\widehat{\phi}(s)$ is the Laplace transform of the self-excitation kernel $\phi(t)$ of the decoder. 

The path-integral for the marginal likelihood (\ref{eq:pathintegral}) is carried out by changing the variable from $\{\nu(t)\}$ to $\{x(t)\}$:
\begin{equation}
p_{\alpha,\gamma}(\{t_i\}) = 
\frac{1}{Z(\gamma)}
\int \exp\left( -\int_0^TL(\dot{x},x)dt \right)D\{x(t)\},
\label{eq:pathintegral_x}
\end{equation}
where the Lagrangian is expressed using Eqs.~(\ref{eq:martingale_decomposition}), (\ref{eq:conditional_intensity_true_decomposition}) and (\ref{eq:conditional_intensity_decoder_decomposition}) as 
\begin{eqnarray}
L(\dot{x},x)
 &=& \frac{1}{2\gamma^2}\dot{x}^2 + 
\Lambda^* + x(t) + \sigma^*Z(t) + \Xi(t) \nonumber\\
& &  - 
\log(\Lambda^* + x(t) + \sigma^*Z(t) + \Xi(t)) \nonumber\\
& &  \times
(\Lambda^* + \sigma^*\zeta(t) + \sigma^* Z^*(t) + \Xi^*(t) 
\nonumber\\
& & 
+ \xi(t) ).
\label{eq:Lagrangian}
\end{eqnarray}
Using this Lagrangian, the path-integral (\ref{eq:pathintegral_x}) is evaluated as Eq.~(\ref{eq:marginal_likelihood_approx}).
We derive the contributions of the MAP solution and the ``quantum effect" to the path-integral below.

%%%%%%%%%%%%%%%%%%%%%%%%%%%%%%%%%%%
\subsection{Contribution of the MAP solution}

Expanding the Lagrangian Eq.~(\ref{eq:Lagrangian}) in terms of $x(t)$, and ignoring $o({\sigma^*}^2/\mu^*)$ and the irrelevant terms yields
%%%
\begin{eqnarray}
\lefteqn{
L(\dot{x},x)
}\hspace{0.1cm}\nonumber\\
 &\approx &
\frac{\dot{x}^2}{2\gamma^2} + \frac{x^2}{2\Lambda^*} 
\nonumber\\
&& - 
\frac{\sigma^* \zeta(t) + \sigma^*Z^*(t)-\sigma^*Z(t) + \Xi^*(t)-\Xi(t) + \xi(t)}{\Lambda^*}x \nonumber\\
&& + \frac{{\sigma^*}^2(2Z^*(t)Z(t)-Z(t)^2)+2\Xi^*(t)\Xi(t)-\Xi(t)^2}{2\Lambda^*} 
\nonumber\\
&& - \frac{{\sigma^*}^2}{\Lambda^*}Z(t)\zeta(t),
\label{eq:Lagrangian_approx}
\end{eqnarray}
for which a solution of the Euler-Lagrange equation (\ref{eq:Euler-Lagrange}) is obtained as
\begin{eqnarray}
\hat{x}(t) &=& \frac{\gamma}{2\sqrt{\Lambda^*}}\int_0^Tdu e^{-\frac{\gamma}{\sqrt{\Lambda^*}}|t-u|} ( \sigma^*\zeta(u) +\sigma^*Z^*(u) \nonumber\\
&& - \sigma^*Z(u) + \Xi^*(u)-\Xi(u) + \xi(u) ).
\label{eq:MAP}
\end{eqnarray}

For the Lagrangian (\ref{eq:Lagrangian_approx}),
the action integral (\ref{eq:action}) is given as
\begin{eqnarray}
\lefteqn{
S[\hat{x}(t)]
}\hspace{0.1cm}\nonumber\\
&=&
\int_0^Tdt \bigg( \frac{1}{2\gamma^2}\frac{d}{dt}(\hat{x}(t)\dot{\hat{x}}(t)) \nonumber\\
&&
- \frac{\sigma^*(\zeta(t)+Z^*(t)-Z(t)) + \Xi^*(t)-\Xi(t) + \xi(t)}{2\Lambda^*}\hat{x}(t) \nonumber\\
&& - 
\frac{{\sigma^*}^2(2Z^*(t)Z(t)-Z(t)^2)+2\Xi^*(t)\Xi(t)-\Xi(t)^2}{2\Lambda^*}\nonumber\\
&&
- \frac{{\sigma^*}^2}{\Lambda^*}Z(t)\zeta(t)
\bigg) .
\end{eqnarray}
Here, the first term on the right-hand-side is a boundary effect, which is negligible in comparison to the bulk contribution given the long event series $T \gg 1$.
Substituting Eq.~(\ref{eq:MAP}) into this formula and assuming ergodicity, we obtain the contribution of the MAP solution as 
\begin{eqnarray}
\lefteqn{
S[\hat{x}(t)]
}\hspace{0.5cm}\nonumber\\
&=&
-\frac{\gamma}{4\Lambda^*\sqrt{\Lambda^*}}\int_0^Tdt\int_0^Tdue^{-\frac{\gamma}{\sqrt{\Lambda^*}}|t-u|}
\nonumber\\
&& \times
\big\{ \langle \xi(t)\xi(u)\rangle 
+ 2\langle \Xi^*(t)\xi(u) \rangle - 2\langle \Xi(t)\xi(u) \rangle 
\nonumber\\
&& 
+\langle \Xi^*(t)\Xi^*(u) \rangle - 2 \langle \Xi^*(t)\Xi(u) \rangle + \langle \Xi(t)\Xi(u) \rangle 
\nonumber\\
&&
+ {\sigma^*}^2 ( \langle \zeta(t)\zeta(u) \rangle 
+ 2\langle Z^*(t)\zeta(u)\rangle - 2\langle Z(t)\zeta(u) \rangle \nonumber\\
&& 
+ \langle Z^*(t)Z^*(u)\rangle -2\langle Z^*(t)Z(u)\rangle + \langle Z(t)Z(u)\rangle ) \big\}
\nonumber\\
& & { } 
- \frac{1}{2\Lambda^*}\left(
2\int_0^Tdt\langle \Xi^*(t)\Xi(t) \rangle - \int_0^Tdt\langle \Xi(t)^2 \rangle \right)
\nonumber\\
& & { } - \frac{{\sigma^*}^2}{2\Lambda^*} \left( 2\int_0^Tdt\langle Z^*(t)Z(t) \rangle - \int_0^Tdt\langle Z(t)^2 \rangle \right) \nonumber\\
& & { } -
\frac{{\sigma^*}^2}{\Lambda^*}\int_0^Tdt\langle Z(t)\zeta(t) \rangle .
\label{eq:classical_contribution}
\end{eqnarray}

%%%%%%%%%%%%%%%%%%%%%%%%%%%%%%%%%%%
\subsection{Contribution of the quantum effect}

The quantum effect (\ref{eq:quantum_effect}) is given by the ratio of functional determinants of the second order differential operators associated with the Lagrangian (\ref{eq:Lagrangian_approx}) \cite{Coleman1988,Kleinert2009},
\begin{equation}
R =
\left[
\frac{\det\big( -\frac{1}{\gamma^2}\partial^2_t + \frac{1}{\Lambda^*} \big) }
{\det\big( -\frac{1}{\gamma^2}\partial^2_t \big) }
\right]^{-\frac{1}{2}}.
\end{equation}
The Gelfand--Yaglom method allows us to calculate the functional determinants by an initial-value problem for the corresponding differential operator \cite{Gelfand1960},
\begin{eqnarray*}
-(1/\gamma^2) \ddot{\psi}_1(t) + \psi_1(t)/\Lambda^* &=& 0,
\quad \psi_1(0)=0,\ \dot{\psi}_1(0)=1, \\
-(1/\gamma^2) \ddot{\psi}_2(t) &=& 0, 
\quad \psi_2(0)=0,\ \dot{\psi}_2(0)=1.
\end{eqnarray*}
Then, the Gelfand--Yaglom reads 
\begin{equation}
\left[
\frac{\det\big( -\frac{1}{\gamma^2}\partial^2_t + \frac{1}{\Lambda^*} \big) }
{\det\big( -\frac{1}{\gamma^2}\partial^2_t \big) }
\right]^{-\frac{1}{2}} =
\left[
\frac{\psi_1(T)}{\psi_2(T)}
\right]^{-\frac{1}{2}} .
\end{equation}
By solving the differential equations, the quantum contribution is obtained as 
\begin{equation}
R = \left( \frac{\sqrt{\Lambda^*}}{2\gamma T} \right)^{-\frac{1}{2}}
\exp\left( -\frac{\gamma T}{2\sqrt{\Lambda^*}}\right). 
\label{eq:quantum_effect_solution}
\end{equation}

%%%%%%%%%%%%%%%%%%%%%%%%%%%%%%%%%%%
\subsection{Formula for free energy}
The free-energy formula is obtained by substituting Eqs.~(\ref{eq:classical_contribution}) and (\ref{eq:quantum_effect_solution}) into Eq.~(\ref{eq:free_energy_def}) as, 
\begin{eqnarray}
\lefteqn{
F(\alpha,\gamma)
}\hspace{0.1cm}\nonumber\\
&=&
\frac{\gamma}{4\sqrt{\Lambda^*}} - \frac{{\sigma^*}^2}{2\Lambda^*}\widetilde{\rho}(0)
- \frac{\widetilde{\Psi}^*(0)-\widetilde{\Psi}(0)}{2} \nonumber\\
&& - 
\frac{1}{2}\bigg(
\int_0^{\infty}dt(2\Psi^*(t)-\Psi(t))\Psi(t) \nonumber\\
&& + 
\int_0^{\infty}dt(\Psi^*(t)-\Psi(t))(\widetilde{\Psi}^*(t)-\widetilde{\Psi}(t))
\bigg) \nonumber\\
&&  -
\frac{{\sigma^*}^2}{\Lambda^*}
\bigg( \int_0^{\infty}dt\Psi(t)\rho(t)
+ \int_0^{\infty}dt(\Psi^*(t)-\Psi(t))\widetilde{\rho}(t) 
\bigg) \nonumber\\
&&  -
\frac{{\sigma^*}^2}{2\Lambda^*}
\bigg(
\int_0^{\infty}dt\int_0^{\infty}du (2\Psi^*(t)-\Psi(t))\Psi(u)\rho(t-u) \nonumber\\
&& +
\int_0^{\infty}dt\int_0^{\infty}du (\Psi^*(t)-\Psi(t))(\Psi^*(u)-\Psi(u))\nonumber\\
&&\times \widetilde{\rho}(t-u)
\bigg) ,
\label{eq:free_energy_general}
\end{eqnarray}
where 
$\rho(u) = \langle \zeta(t)\zeta(t+u) \rangle$ is the normalized autocorrelation function of the original exogenous input, and 
\begin{eqnarray}
\widetilde{\rho}(t) &=& \frac{\gamma}{2\sqrt{\Lambda^*}}\int_{-\infty}^{\infty}du e^{-\frac{\gamma}{\sqrt{\Lambda^*}}|u|}\rho(t+u), \\
\widetilde{\Psi}(t) &=& \frac{\gamma}{\sqrt{\Lambda^*}}\int_0^{\infty}du e^{-\frac{\gamma}{\sqrt{\Lambda^*}}u}\Psi(t+u), \\
\widetilde{\Psi}^*(t) &=& \frac{\gamma}{\sqrt{\Lambda^*}}\int_0^{\infty}du e^{-\frac{\gamma}{\sqrt{\Lambda^*}}u}\Psi^*(t+u).
\end{eqnarray}

For the exponential kernels $\phi^*(t)=1/\tau^*_s\exp(-t/\tau^*_s)$ and $\phi(t)=1/\tau_s\exp(-t/\tau_s)$, the effective self-excitation kernels are, respectively, obtained as 
\begin{equation}
\Psi^*(t) = \alpha^*/\tau^*_se^{-(1-\alpha^*)t/\tau^*_s}, 
\label{eq:effective_kernel_ture_exp}
\end{equation}
and 
%%%%
\begin{eqnarray}
\Psi(t) &=& \frac{\alpha\alpha^*}{\tau_s^*-(1-\alpha^*)\tau_s}e^{-(1-\alpha^*)t/\tau^*_s} \nonumber\\
&& + \frac{\alpha(\tau^*_s/\tau_s-1)}{\tau^*_s-(1-\alpha^*)\tau_s}e^{-t/\tau_s},
\label{eq:effective_kernel_model_exp}
\end{eqnarray}
for $\tau_s \neq \tau^*_s/(1-\alpha^*)$, and 
\begin{equation}
\Psi(t) = \alpha(1-\alpha^*)(1+\alpha^*t/\tau^*_s)/\tau^*_s e^{-(1-\alpha^*)t/\tau^*_s},
\end{equation}
for $\tau_s = \tau^*_s/(1-\alpha^*)$. 
Substituting Eqs.~(\ref{eq:effective_kernel_ture_exp}) and (\ref{eq:effective_kernel_model_exp}) and the autocorrelation function of the OUP, $\rho(u) = \exp(-|u|/\tau^*_e)$, into Eq.~(\ref{eq:free_energy_general}) yields the free energy for the Hawkes decoder with the exponential kernels. 
By scaling the hyperparameter $\gamma \leftarrow \gamma\tau_e^*/\sqrt{\Lambda^*}$ and using the relative time constants, $\beta_s=\tau_e^*/\tau_s$ and $\beta^*_s=\tau_e^*/\tau^*_s$, the dimensionless free energy, $F \leftarrow \tau_e^* F$, is expressed as 
\begin{eqnarray}
\lefteqn{
F(\alpha,\gamma)
}\hspace{0.1cm} \nonumber\\
&=& 
%%% 1 %%%
\frac{\gamma}{4} - \frac{{\sigma^*}^2\tau^*_e}{\mu^*} 
\frac{(1-\alpha^*)\gamma}{2(\gamma+1)} 
- \frac{\gamma}{2}\bigg\{ \frac{c_0-c_1(\alpha)}{\gamma+\beta_{\alpha}^*} - \frac{c_2(\alpha)}{\gamma+\beta_s} \bigg\} 
\nonumber\\
&& -
%%% 2 %%%
\frac{1}{2}\bigg\{
\frac{(c_0-c_1(\alpha))^2\gamma}{2\beta_{\alpha}^*(\gamma+\beta_{\alpha}^*)}
- \frac{(c_0-c_1(\alpha))c_2(\alpha)\gamma}{(\beta_{\alpha}^*+\beta_s)(\gamma+\beta_s)} \nonumber\\
&&
- \frac{(c_0-c_1(\alpha))c_2(\alpha)\gamma}{(\beta_{\alpha}^*+\beta_s)(\gamma+\beta_{\alpha}^*)}
+ \frac{c_2(\alpha)^2\gamma}{2\beta_s(\gamma+\beta_s)} 
- \frac{c_2(\alpha)^2}{2\beta_s}
\nonumber\\
&& + 
\frac{(2c_0-c_1(\alpha))c_1(\alpha)}{2\beta_{\alpha}^*} + \frac{2(c_0-c_1(\alpha))c_2(\alpha)}{\beta_{\alpha}^*+\beta_s} 
\bigg\}
\nonumber\\
&& - 
%%% 3 %%%
\frac{{\sigma^*}^2\tau^*_e}{\mu^*} (1-\alpha^*)\bigg\{
\frac{(c_0-c_1(\alpha))(\gamma+\beta_{\alpha}^*+1)\gamma}{(\beta_{\alpha}^*+1)(\gamma+\beta_{\alpha}^*)(\gamma+1)} \nonumber\\
&&
- \frac{c_2(\alpha)(\gamma+\beta_s+1)\gamma}{(\beta_s+1)(\gamma+\beta_s)(\gamma+1)} + 
\frac{c_1(\alpha)}{\beta_{\alpha}^*+1} + \frac{c_2(\alpha)}{\beta_s+1}
\bigg\} \nonumber\\
&& - 
%%% 4 %%%
\frac{{\sigma^*}^2\tau^*_e}{\mu^*}
\frac{(1-\alpha^*)}{2}\bigg\{
\frac{(c_0-c_1(\alpha))^2(\gamma+\beta_{\alpha}^*+1)\gamma}{\beta_{\alpha}^*(\beta_{\alpha}^*+1)(\gamma+\beta_{\alpha}^*)(\gamma+1)} \nonumber\\
&&
- \frac{2(c_0-c_1(\alpha))c_2(\alpha)(\gamma+\beta_{\alpha}^*+1)\gamma}{(\beta_{\alpha}^*+\beta_s)(\beta_{\alpha}^*+1)(\gamma+\beta_{\alpha}^*)(\gamma+1)} \nonumber\\
&& -
\frac{2(c_0-c_1(\alpha))c_2(\alpha)(\gamma+\beta_s+1)\gamma}{(\beta_{\alpha}^*+\beta_s)(\beta_s+1)(\gamma+\beta_s)(\gamma+1)} \nonumber\\
&&
+ \frac{c_2(\alpha)^2(\gamma+\beta_s+1)\gamma}{\beta_s(\beta_s+1)(\gamma+\beta_s)(\gamma+1)} 
- \frac{c_2(\alpha)^2}{\beta_s(\beta_s+1)}
\nonumber\\
&&
+ \frac{(2c_0-c_1(\alpha))c_1(\alpha)}{\beta_{\alpha}^*(\beta_{\alpha}^*+1)} 
+ \frac{2(c_0-c_1(\alpha))c_2(\alpha)}{(\beta_{\alpha}^*+\beta_s)(\beta_s+1)}
 \nonumber\\
&&
+ \frac{2(c_0-c_1(\alpha))c_2(\alpha)}{(\beta_{\alpha}^*+\beta_s)(\beta_{\alpha}^*+1)}
\bigg\},
\label{eq:free_energy_oup}
\end{eqnarray}
where
$\beta_{\alpha}^* = (1-\alpha^*)\beta^*_s$, $c_0 = \alpha^*\beta^*_s$, 
$c_1(\alpha) = \alpha\alpha^*\beta_s\beta^*_s/(\beta_s-\beta^*_{\alpha})$
and 
$c_2(\alpha) = \alpha\beta_s(\beta_s-\beta^*_s)/(\beta_s-\beta^*_{\alpha})$.
Note that the free energy (\ref{eq:free_energy_oup}) is fully characterized by the four dimensionless parameters: 
the strengths of endogenous self-excitation $\alpha^*$ and the exogenous fluctuations ${\sigma^*}^2\tau^*_e/\mu^*$, and the time constants of the self-excitation kernels relative to that of original exogenous input, $\tau^*_s/\tau^*_e (= {\beta^*_s}^{-1})$ and $\tau_s/\tau^*_e (=\beta_s^{-1})$.

%%%%%%%%%%%%%%%%%%%%%%%%%%%%%%%%%%%
\section{Analysis of free energy}
%%%%%%%%%%%%%%%%%%%%%%%%%%%%%%%%%%%
\subsection{The Hawkes decoder with a correct self-excitation kernel}
\label{appendix:correct_self-excitation_kernel}
%%%%%%%%%%%%%%%%%%%%%%%%%%%%%%%%%%%
Here, we analyze the free energy (\ref{eq:free_energy_oup}) in the case where the model uses the correct shape of the original self-excitation kernel, $\tau_s = \tau^*_s$. 
In particular, we consider the case of slow exogenous fluctuation, $\tau^*_e \gg \tau^*_s$. 
The free energy (\ref{eq:free_energy_oup}) is asymptotically evaluated as 
\begin{eqnarray}
\lefteqn{
F(\alpha,\gamma)
}\hspace{0.1cm}\nonumber\\
 &=& 
-\frac{\alpha(2\alpha^*-\alpha)}{4(1-\alpha^*)}\frac{\tau^*_e}{\tau^*_s}
+ \bigg\{
\frac{\gamma}{4} - \frac{{\sigma^*}^2\tau^*_e}{\mu^*}\frac{(1-\alpha^*)\gamma}{2(\gamma+1)} \nonumber\\
&&
-\frac{(\alpha^*-\alpha)\gamma}{2(1-\alpha^*)}
-\frac{(\alpha^*-\alpha)^2\gamma}{4(1-\alpha^*)^2}
+ \frac{{\sigma^*}^2\tau^*_e}{\mu^*}\frac{(\alpha^*\gamma+\alpha)}{\gamma+1} \nonumber\\
&&
- \frac{{\sigma^*}^2\tau^*_e}{\mu^*}\frac{({\alpha^*}^2\gamma+2\alpha\alpha^*-\alpha^2)}{2(1-\alpha^*)(\gamma+1)}
\bigg\} + O(\tau^*_s/\tau^*_e),
\end{eqnarray}
from which we obtain $\hat{\alpha}=\alpha^*+O(\tau^*_s/\tau^*_e)$, i.e., the reproduction ratio can be estimated with a potential small error of the order of $\tau^*_s/\tau^*_e$.
In the limit of $\tau^*_s/\tau^*_e \ll 1$, the free energy evaluated at $\alpha=\hat{\alpha}$ is given as a function of $\gamma$ as
\begin{equation}
F(\hat{\alpha},\gamma) = \frac{\gamma}{4} - \frac{{\sigma^*}^2\tau^*_e}{\mu^*}\frac{(1-\alpha^*)\gamma}{2(\gamma+1)},
\end{equation}
which has a minimum at $\gamma=0$ if 
\begin{equation}
{\sigma^*}^2\tau^*_e/\mu^* < \frac{1}{2(1-\alpha^*)}.
\label{eq:condition_finite_gamma}
\end{equation}

%%%%%%%%%%%%%%%%%%%%%%%%%%%%%%%%%%%
\subsection{The Bayesian rate estimator}
\label{appendix:bayesian_rate_estimator}
%%%%%%%%%%%%%%%%%%%%%%%%%%%%%%%%%%%
The free energy for the Bayesian rate estimator is obtained by setting $\alpha=0$ in Eq.~(\ref{eq:free_energy_oup}) as 
\begin{eqnarray}
\lefteqn{
F_{\rm p}(\gamma) = F(\alpha=0,\gamma)
}\hspace{0.1cm}\nonumber\\
&=&
\frac{\gamma}{4} 
- \frac{\alpha^*(2-\alpha^*)\beta^*_s\gamma}{4(1-\alpha^*)(\gamma+(1-\alpha^*)\beta^*_s)}
\nonumber\\
&&
- \frac{{\sigma^*}^2\tau^*_e}{\mu^*} \frac{\alpha^*(2-\alpha^*)\beta^*_s(1+\gamma+(1-\alpha^*)\beta^*_s)\gamma}{2(\gamma+(1-\alpha^*)\beta^*_s)(1+(1-\alpha^*)\beta^*_s)(\gamma+1)} \nonumber\\
&& 
- \frac{{\sigma^*}^2\tau^*_e}{\mu^*}\frac{(1-\alpha^*)\gamma}{2(\gamma+1)} .
\label{eq:free_energy_rate_estimator}
\end{eqnarray}
The conditions under which the free energy (\ref{eq:free_energy_rate_estimator}) has a minimum at $\gamma \neq 0$ are analytically derived into two particular cases. 
For the case of $\alpha^*=0$, in which data is generated by the (inhomogeneous) Poisson process, Eq.~(\ref{eq:free_energy_rate_estimator}) becomes
\begin{equation}
F_{\rm p}(\gamma) = \frac{\gamma}{4} - \frac{{\sigma^*}^2\tau^*_e}{\mu^*}\frac{\gamma}{2(\gamma+1)},
\end{equation}
which corresponds to the free energy derived in \cite{Koyama2007}.
The condition for $\hat{\gamma} \neq 0$ is then obtained as 
$ {\sigma^*}^2\tau^*_e/\mu^* > 1/2$. 

For the other case, in which the data is generated by the homogenous Hawkes process (${\sigma^*}^2\tau^*_e/\mu^*=0$), the free energy (\ref{eq:free_energy_rate_estimator}) becomes
\begin{eqnarray}
F_{\rm p}(\gamma) &=&
\frac{\gamma((1-\alpha^*)\gamma+(2{\alpha^*}^2-4\alpha^*+1)\beta^*_s)}{4(1-\alpha^*)(\gamma+(1-\alpha^*)\beta^*_s)}, 
\end{eqnarray}
from which the condition for $\hat{\gamma} \neq 0$ is derived as $\alpha^* > \alpha_c = 1- 1/\sqrt{2}$.

%%%%%%%%%%%%%%%%%%%%%%%%%%%%%%%%%%%
\subsection{The Hawkes decoder with an incorrect self-excitation kernel}
\label{appendix:incorrect_self-excitation_kernel}
%%%%%%%%%%%%%%%%%%%%%%%%%%%%%%%%%%%
We analyze the free energy for the Hawkes decoder with an incorrect self-excitation kernel ($\tau_s \neq \tau^*_s$) in the limit of $\tau_s/\tau^*_e \ll 1$ while $\tau^*_s/\tau^*_e$ remains finite (which includes the cases of $\tau^*_e \gg \tau^*_s$ and $\tau_s \ll \tau^*_s$).
The free energy (\ref{eq:free_energy_oup}) is asymptotically expanded with respect to $\tau_s/\tau^*_e$ as 
\begin{equation}
F(\alpha,\gamma) = \frac{\alpha^2}{4}\frac{\tau^*_e}{\tau_s} + 
F_{\rm p}(\alpha,\gamma) + O(\tau_s/\tau^*_e),
\label{eq:free_energy_wrong_kernel_1}
\end{equation}
where $F_{\rm p}(\alpha,\gamma)$ represents a collection of constant terms that satisfy $F_{\rm p}(\alpha=0,\gamma)=F_{\rm p}(\gamma)$.
From Eq.~(\ref{eq:free_energy_wrong_kernel_1}) we obtain $\hat{\alpha}=O(\tau_s/\tau^*_e)$, i.e., the estimated reproduction ratio becomes infinitesimal. In this limit, the free energy is identical to that of the Bayesian rate decoder, $F_{\rm p}(\gamma)$.

%%% Appendix %%%%%%%%%%%%%%%%%%%%%%%%%%%%%%%%%%%%%%%%%%%%%%%%%%%%%%%%%%%%%%%%%%
%%%%%%%%%%%%%%%%%%%%%%%%%%%%%%%%%%%%%%%%%%%%%%%%%%%%%%%%%%%%%%%%%%%%%%%%%%%%%

\section*{ACKNOWLEDGMENTS}
We thank Hidetaka Manabe for his technical assistance in developing a web-application program. S.S. is supported by JSPS KAKENHI Grant number 26280007, JST CREST Grant Number JPMJCR1304, and the New Energy and Industrial Technology Development Organization (NEDO).

%%% References %%%%%%%%%%%%%%%%%%%%%%%%%%%%%%%%%
%%%%%%%%%%%%%%%%%%%%%%%%%%%%%%%%%%%%%%%%%


\begin{thebibliography}{10}

\bibitem{Hethcote2000}
H.~W. Hethcote.
\newblock The mathematics of infectious diseases.
\newblock {\em SIAM review}, Vol.~42, No.~4, pp. 599--653, 2000.

\bibitem{Keeling2011}
M.~J. Keeling and P.~Rohani.
\newblock {\em Modeling infectious diseases in humans and animals}.
\newblock Princeton University Press, 2011.

\bibitem{Vespignani2012}
A.~Vespignani.
\newblock Modelling dynamical processes in complex socio-technical systems.
\newblock {\em Nature physics}, Vol.~8, No.~1, p.~32, 2012.

\bibitem{Brockmann2013}
D.~Brockmann and D.~Helbing.
\newblock The hidden geometry of complex, network-driven contagion phenomena.
\newblock {\em Science}, Vol. 342, No. 6164, pp. 1337--1342, 2013.

\bibitem{Pastorsatorras2015}
R.~Pastor-Satorras, C.~Castellano, P.~Van Mieghem, and A.~Vespignani.
\newblock Epidemic processes in complex networks.
\newblock {\em Reviews of Modern Physics}, Vol.~87, No.~3, p. 925, 2015.

\bibitem{Goffman1964}
W.~Goffman and V.~A. Newill.
\newblock Generalization of epidemic theory.
\newblock {\em Nature}, Vol. 204, No. 4955, pp. 225--228, 1964.

\bibitem{Kitsak2010}
M.~Kitsak, L.~K. Gallos, S.~Havlin, F.~Liljeros, L.~Muchnik, H.~E. Stanley, and
  H.~A. Makse.
\newblock Identification of influential spreaders in complex networks.
\newblock {\em Nature physics}, Vol.~6, No.~11, p. 888, 2010.

\bibitem{Mohler2011}
G.~O. Mohler, M.~B. Short, P.~J. Brantingham, F.~P. Schoenberg, and G.~E. Tita.
\newblock Self-exciting point process modeling of crime.
\newblock {\em Journal of the American Statistical Association}, Vol. 106, No.
  493, pp. 100--108, 2011.

\bibitem{Lewis2012}
E.~Lewis, G.~Mohler, P.~J. Brantingham, and A.~L. Bertozzi.
\newblock Self-exciting point process models of civilian deaths in iraq.
\newblock {\em Security Journal}, Vol.~25, No.~3, pp. 244--264, Jul 2012.

\bibitem{ogata1988}
Y.~Ogata.
\newblock Statistical models for earthquake occurrences and residual analysis
  for point processes.
\newblock {\em Journal of the American Statistical Association}, Vol.~83, No.
  401, pp. 9--27, 1988.

\bibitem{helmstetter2003}
A.~Helmstetter and D.~Sornette.
\newblock Predictability in the epidemic-type aftershock sequence model of
  interacting triggered seismicity.
\newblock {\em Journal of Geophysical Research: Solid Earth}, Vol. 108, No.
  B10, p. 2482, 2003.

\bibitem{michael2012}
M.~Golosovsky and S.~Solomon.
\newblock Stochastic dynamical model of a growing citation network based on a
  self-exciting point process.
\newblock {\em Physical Review Letters}, Vol. 109, p. 098701, Aug 2012.

\bibitem{CraneSornette2008}
R.~Crane and D.~Sornette.
\newblock Robust dynamic classes revealed by measuring the response function of
  a social system.
\newblock {\em Proceedings of the National Academy of Sciences}, Vol. 105,
  No.~41, pp. 15649--15653, 2008.

\bibitem{Lerman2010}
K.~Lerman and R.~Ghosh.
\newblock Information contagion: An empirical study of the spread of news on
  digg and twitter social networks.
\newblock {\em ICWSM}, Vol.~10, pp. 90--97, 2010.

\bibitem{Romero2011}
D.~M. Romero, B.~Meeder, and J.~Kleinberg.
\newblock Differences in the mechanics of information diffusion across topics:
  idioms, political hashtags, and complex contagion on twitter.
\newblock pp. 695--704. ACM, 2011.

\bibitem{AtSahalia2010}
Y.~A{\"\i}t-Sahalia, J.~Cacho-Diaz, and R.~J.~A. Laeven.
\newblock Modeling financial contagion using mutually exciting jump processes.
\newblock Technical report, 2010.

\bibitem{Hardiman2013}
S.~J. Hardiman, N.~Bercot, and J.~P. Bouchaud.
\newblock Critical reflexivity in financial markets: a {Hawkes} process
  analysis.
\newblock {\em The European Physical Journal B}, Vol.~86, No.~10, p. 442, 2013.

\bibitem{bacry2015}
E.~Bacry, I.~Mastromatteo, and J.~F. Muzy.
\newblock {Hawkes} processes in finance.
\newblock {\em Market Microstructure and Liquidity}, Vol.~1, p. 1550005, 2015.

\bibitem{hawkes2018}
A.~G. Hawkes.
\newblock Hawkes processes and their applications to finance: a review.
\newblock {\em Quantitative Finance}, Vol.~18, No.~2, pp. 193--198, 2018.

\bibitem{Pernice2011}
V.~Pernice, B.~Staude, S.~Cardanobile, and S.~Rotter.
\newblock How structure determines correlations in neuronal networks.
\newblock {\em PLoS Computational Biology}, Vol.~7, No.~5, p. e1002059, 2011.

\bibitem{ReynaudBouret2014}
P.~Reynaud-Bouret, V.~Rivoirard, F.~Grammont, and C.~Tuleau-Malot.
\newblock Goodness-of-fit tests and nonparametric adaptive estimation for spike
  train analysis.
\newblock {\em The Journal of Mathematical Neuroscience}, Vol.~4, No.~1, p.~3,
  2014.

\bibitem{Kass2018}
R.~E. Kass, S.~Amari, K.~Arai, E.~N. Brown, C.~O. Diekman, M.~Diesmann,
  B.~Doiron, U.~T. Eden, A.~L. Fairhall, G.~M. Fiddyment, T.~Fukai,
  S.~Gr\"{u}n, M.~T. Harrison, M.~Helias, H.~Nakahara, J.~Teramae, P.~J.
  Thomas, M.~Reimers, J.~Rodu, H.~G. Rotstein, E.~Shea-Brown, H.~Shimazaki,
  S.~Shinomoto, B.~M. Yu, and M.~A. Kramer.
\newblock Computational neuroscience: Mathematical and statistical
  perspectives.
\newblock {\em Annual Review of Statistics and Its Application}, Vol.~5, No.~1,
  pp. 183--214, 2018.

\bibitem{GerhardDegerTruccolo2017}
F.~Gerhard, M.~Deger, and W.~Truccolo.
\newblock On the stability and dynamics of stochastic spiking neuron models:
  Nonlinear {Hawkes} process and point process {GLM}s.
\newblock {\em PLoS Computational Biology}, Vol.~13, No.~2, p. e1005390, 2017.

\bibitem{Lipp2002}
E.~K. Lipp, A.~Huq, and R.~R. Colwell.
\newblock Effects of global climate on infectious disease: the cholera model.
\newblock {\em Clinical Microbiology Reviews}, Vol.~15, No.~4, pp. 757--770,
  2002.

\bibitem{Menezes2004}
M.~Argollo de~Menezes and A.~L. Barab\'asi.
\newblock Fluctuations in network dynamics.
\newblock {\em Physical Review Letters}, Vol.~92, No.~2, p. 028701, 2004.

\bibitem{lazer2018science}
D.~M.~J. Lazer, M.~A. Baum, Y.~Benkler, A.~J. Berinsky, K.~M. Greenhill,
  F.~Menczer, M.~J. Metzger, B.~Nyhan, G.~Pennycook, D.~Rothschild, et~al.
\newblock The science of fake news.
\newblock {\em Science}, Vol. 359, No. 6380, pp. 1094--1096, 2018.

\bibitem{Omi2017}
T.~Omi, Y.~Hirata, and K.~Aihara.
\newblock Hawkes process model with a time-dependent background rate and its
  application to high-frequency financial data.
\newblock {\em Physical Review E}, Vol.~96, No.~1, p. 012303, 2017.

\bibitem{zaman2014}
T.~Zaman, E.~B. Fox, and E.~T. Bradlow.
\newblock A {Bayesian} approach for predicting the popularity of tweets.
\newblock {\em The Annals of Applied Statistics}, Vol.~8, No.~3, pp.
  1583--1611, 2014.

\bibitem{cheng2014}
J.~Cheng, L.~Adamic, P.~A. Dow, J.~M. Kleinberg, and J.~Leskovec.
\newblock Can cascades be predicted?
\newblock In {\em WWW' 14}, pp. 925--936, 2014.

\bibitem{Dow2013}
P.~A. Dow, L.~A. Adamic, and A.~Friggeri.
\newblock The anatomy of large facebook cascades.
\newblock In {\em ICWSM' 13}, pp. 145--154, 2013.

\bibitem{petrovic2011rt}
S.~Petrovic, M.~Osborne, and V.~Lavrenko.
\newblock Rt to win! predicting message propagation in twitter.
\newblock In {\em ICWSM' 11}, pp. 586--589, 2011.

\bibitem{Aoki2016}
T.~Aoki, T.~Takaguchi, R.~Kobayashi, and R.~Lambiotte.
\newblock Input-output relationship in social communications characterized by
  spike train analysis.
\newblock {\em Physical Review E}, Vol.~94, No.~4, p. 042313, 2016.

\bibitem{Cattuto2007}
C.~Cattuto, V.~Loreto, and L.~Pietronero.
\newblock Semiotic dynamics and collaborative tagging.
\newblock {\em Proceedings of the National Academy of Sciences}, Vol. 104,
  No.~5, pp. 1461--1464, 2007.

\bibitem{Rybski2009}
D.~Rybski, S.~V. Buldyrev, S.~Havlin, F.~Liljeros, and H.~A. Makse.
\newblock Scaling laws of human interaction activity.
\newblock {\em Proceedings of the National Academy of Sciences}, Vol. 106,
  No.~31, pp. 12640--12645, 2009.

\bibitem{Fujita2018}
K.~Fujita, A.~Medvedev, S.~Koyama, R.~Lambiotte, and S.~Shinomoto.
\newblock Identifying exogenous and endogenous activity in social media.
\newblock {\em Phys. Rev. E}, Vol.~98, p. 052304, Nov 2018.

\bibitem{hawkes1971}
A.~G. Hawkes.
\newblock Spectra of some self-exciting and mutually exciting point processes.
\newblock {\em Biometrika}, Vol.~58, No.~1, pp. 83--90, 1971.

\bibitem{Snyder1991}
D.~L. Snyder and M.~I. Miller.
\newblock {\em Random Point Processes in Time and Space}.
\newblock Springer, 2nd edition, 1991.

\bibitem{Daley2002}
D.~Daley and D.~Vere-Jones.
\newblock {\em An Introduction to the Theory of Point Processes}, Vol.~1.
\newblock Springer, 2nd edition, 2002.

\bibitem{Koyama2004}
S.~Koyama and S.~Shinomoto.
\newblock Histogram bin width selection for time-dependent poisson processes.
\newblock {\em Journal of Physics A: Mathematical and General}, Vol.~37,
  No.~29, pp. 7255--7265, jul 2004.

\bibitem{Koyama2007}
S.~Koyama, T.~Shimokawa, and S.~Shinomoto.
\newblock Phase transitions in the estimation of event rate: a path integral
  analysis.
\newblock {\em Journal of Physics A: Mathematical and Theoretical}, Vol.~40,
  No.~20, pp. F383--F390, apr 2007.

\bibitem{Onaga2014}
T.~Onaga and S.~Shinomoto.
\newblock Bursting transition in a linear self-exciting point process.
\newblock {\em Phys. Rev. E}, Vol.~89, p. 042817, Apr 2014.

\bibitem{Onaga2016}
T.~Onaga and S.~Shinomoto.
\newblock Emergence of event cascades in inhomogeneous networks.
\newblock {\em Scientific Reports}, Vol.~6, pp. 33321 EP --, Sep 2016.
\newblock Article.

\bibitem{Ogata1978}
Y.~Ogata.
\newblock The asymptotic behavior of maximum likelihood estimators for
  stationary point processes.
\newblock {\em Annals of the Institute of Statistical Mathematics}, Vol.~30,
  No.~2, pp. 243--261, 1978.

\bibitem{Ozaki1979}
T.~Ozaki.
\newblock Maximum likelihood estimation of {Hawkes}' self-exciting point
  processes.
\newblock {\em Annals of the Institute of Statistical Mathematics}, Vol.~31,
  No.~1, pp. 145--155, 1979.

\bibitem{Veen2008}
A.~Veen and F.~P. Schoenberg.
\newblock Estimation of space-time branching process models in seismology using
  an {EM}-type algorithm.
\newblock {\em Journal of the American Statistical Association}, Vol. 103, No.
  482, pp. 614--624, 2008.

\bibitem{Rasmussen2013}
J.~G. Rasmussen.
\newblock Bayesian inference for {Hawkes} processes.
\newblock {\em Methodology and Computing in Applied Probability}, Vol.~15,
  No.~3, pp. 623--642, Sep 2013.

\bibitem{Fonseca2014}
J.~Da Fonseca and R.~Zaatour.
\newblock {Hawkes} process: Fast calibration, application to trade clustering,
  and diffusive limit.
\newblock {\em Journal of Futures Markets}, Vol.~34, No.~6, pp. 548--579, 2014.

\bibitem{Chen2018}
J.~Chen, A.~G. Hawkes, E.~Scalas, and M.~Trinh.
\newblock Performance of information criteria for selection of {Hawkes} process
  models of financial data.
\newblock {\em Quantitative Finance}, Vol.~18, No.~2, pp. 225--235, 2018.

\bibitem{Lewis2011}
E.~Lewis and G.~Mohler.
\newblock A nonparametric {EM} algorithm for multiscale {Hawkes} processes.
\newblock preprint, 2011.

\bibitem{Bacry2012}
E.~Bacry, K.~Dayri, and J.~F. Muzy.
\newblock Non-parametric kernel estimation for symmetric {Hawkes} processes.
  application to high frequency financial data.
\newblock {\em The European Physical Journal B}, Vol.~85, p. 157, 2012.

\bibitem{Bacry2016a}
E.~Bacry and J.~F. Muzy.
\newblock First- and second-order statistics characterization of {Hawkes}
  processes and non-parametric estimation.
\newblock {\em IEEE Transactions on Information Theory}, Vol.~62, No.~4, pp.
  2184--2202, April 2016.

\bibitem{Bacry2016b}
E.~Bacry, T.~Jaisson, and J.~F. Muzy.
\newblock Estimation of slowly decreasing {Hawkes} kernels: application to
  high-frequency order book dynamics.
\newblock {\em Quantitative Finance}, Vol.~16, No.~8, pp. 1179--1201, 2016.

\bibitem{Patricia2010}
P.~Reynaud-Bouret and S.~Schbath.
\newblock Adaptive estimation for {Hawkes} processes; application to genome
  analysis.
\newblock {\em The Annals of Statistics}, Vol.~38, No.~5, pp. 2781--2822, 2010.

\bibitem{Hansen2015}
N.~R. Hansen, P.~Reynaud-Bouret, and V.~Rivoirard.
\newblock Lasso and probabilistic inequalities for multivariate point
  processes.
\newblock {\em Bernoulli}, Vol.~21, No.~1, pp. 83--143, 2015.

\bibitem{zhou13}
K.~Zhou, H.~Zha, and L.~Song.
\newblock Learning triggering kernels for multi-dimensional {Hawkes} processes.
\newblock In Sanjoy Dasgupta and David McAllester, editors, {\em Proceedings of
  the 30th International Conference on Machine Learning}, Vol.~28 of {\em
  Proceedings of Machine Learning Research}, pp. 1301--1309, Atlanta, Georgia,
  USA, 17--19 Jun 2013. PMLR.

\bibitem{xuc16}
H.~Xu, M.~Farajtabar, and H.~Zha.
\newblock Learning granger causality for {Hawkes} processes.
\newblock In Maria~Florina Balcan and Kilian~Q. Weinberger, editors, {\em
  Proceedings of The 33rd International Conference on Machine Learning},
  Vol.~48 of {\em Proceedings of Machine Learning Research}, pp. 1717--1726,
  New York, New York, USA, 20--22 Jun 2016. PMLR.

\bibitem{Kirchner2016}
M.~Kirchner.
\newblock {Hawkes} and {INAR}($\infty$) processes.
\newblock {\em Stochastic Processes and their Applications}, Vol. 126, No.~8,
  pp. 2494--2525, 2016.

\bibitem{Kirchner2017}
M.~Kirchner.
\newblock An estimation procedure for the {Hawkes} process.
\newblock {\em Quantitative Finance}, Vol.~17, No.~4, pp. 571--595, 2017.

\bibitem{chen_hall_2016}
F.~Chen and P.~Hall.
\newblock Nonparametric estimation for self-exciting point processes--a
  parsimonious approach.
\newblock {\em Journal of Computational and Graphical Statistics}, Vol.~25,
  No.~1, pp. 209--224, 2016.

\bibitem{Clinet2018}
S.~Clinet and Y.~Potiron.
\newblock Statistical inference for the doubly stochastic self-exciting
  process.
\newblock {\em Bernoulli}, Vol.~24, No.~4B, pp. 3469--3493, 2018.

\bibitem{chen_hall_2013}
F.~Chen and P.~Hall.
\newblock Inference for a nonstationary self-exciting point process with an
  application in ultra-high frequency financial data modeling.
\newblock {\em Journal of Applied Probability}, Vol.~50, No.~4, pp. 1006--1024,
  2013.

\bibitem{kobayashi2016}
R.~Kobayashi and R.~Lambiotte.
\newblock Tideh: Time-dependent {Hawkes} process for predicting retweet
  dynamics.
\newblock In {\em ICWSM' 2016}, pp. 191--200, 2016.

\bibitem{West1997}
M.~West and J.~Harrison.
\newblock {\em Bayesian Forecasting and Dynamic Models}.
\newblock Springer, 2nd edition, 1997.

\bibitem{Durbin2001}
J.~Durbin and S.~Koopman.
\newblock {\em Time series analysis by state space methods}.
\newblock Oxford University Press, 2001.

\bibitem{Koyama2010JASA}
S.~Koyama, L.~C. Perez-Bolde, C.~R. Shalizi, and R.~E. Kass.
\newblock Approximate methods for state-space models.
\newblock {\em Journal of American Statistical Association}, Vol. 105, pp.
  170--180, 2010.

\bibitem{Koyama2010JCNS}
S.~Koyama and L.~Paninski.
\newblock Efficient computation of the maximum a posteriori path and parameter
  estimation in integrate-and-fire and more general state-space models.
\newblock {\em Journal of Computational Neuroscience}, Vol.~29, pp. 89--105,
  2010.

\bibitem{Koyama2010AISM}
S.~Koyama, U.~T. Eden, E.~N. Brown, and R.~E. Kass.
\newblock Bayesian decoding of neural spike trains.
\newblock {\em Annals of the Institute of Statistical Mathematics}, Vol.~62,
  pp. 37--59, 2010.

\bibitem{Press1992}
W.~H. Press, B.~P. Flannery, S.~A. Teukolsky, and W.~T. Vetterling.
\newblock {\em Numerical Recipes in C: The Art of Scientific Computing}.
\newblock Cambridge University Press, 2nd edition, 1992.

\bibitem{Coleman1988}
S.~Coleman.
\newblock {\em Aspects of {Symmetry}}, chapter~7.
\newblock Cambridge University Press, 1988.

\bibitem{Kleinert2009}
H.~Kleinert.
\newblock {\em Path Integrals in Quantum Mechanics, Statistics, Polymer
  Physics, and Financial Markets}.
\newblock World Scientific Publishing Company, 5th edition, 2009.

\bibitem{Gelfand1960}
I.~M. Gelfand and A.~M. Yaglom.
\newblock Integration in functional spaces and applications in quantum physics.
\newblock {\em Journal of Mathematical Physics}, Vol.~1, pp. 48--69, 1960.

\end{thebibliography}
\end{document}